\documentclass[journal]{IEEEtran}

\ifCLASSINFOpdf
\else
\fi
\usepackage{amsmath,amssymb}
\usepackage{algorithm}  
\usepackage{algorithmic}
\usepackage{graphicx,color,rotating}

\DeclareMathOperator*{\argmin}{arg\,min}
\usepackage{cite}
\usepackage{graphicx}
\usepackage{float}
\usepackage{subfigure}
\usepackage{bm}
\usepackage{framed}
\usepackage{makecell}
\usepackage{threeparttable}

\newcommand{\brn}{{\mathbf n}}

\newcommand{\brq}{{\mathbf q}}

\newcommand{\brs}{{\mathbf s}}

\newcommand{\brx}{{\mathbf x}}
\newcommand{\bry}{{\mathbf y}}

\newcommand{\brA}{{\mathbf A}}
\newcommand{\brB}{{\mathbf B}}
\newcommand{\brC}{{\mathbf C}}
\newcommand{\brD}{{\mathbf D}}

\newcommand{\brF}{{\mathbf F}}
\newcommand{\brG}{{\mathbf G}}
\newcommand{\brH}{{\mathbf H}}
\newcommand{\brI}{{\mathbf I}}
\newcommand{\brJ}{{\mathbf J}}
\newcommand{\brK}{{\mathbf K}}
\newcommand{\brL}{{\mathbf L}}
\newcommand{\brM}{{\mathbf M}}

\newcommand{\brO}{{\mathbf O}}
\newcommand{\brP}{{\mathbf P}}
\newcommand{\brQ}{{\mathbf Q}}
\newcommand{\brR}{{\mathbf R}}
\newcommand{\brS}{{\mathbf S}}
\newcommand{\brT}{{\mathbf T}}
\newcommand{\brU}{{\mathbf U}}
\newcommand{\brV}{{\mathbf V}}
\newcommand{\brW}{{\mathbf W}}
\newcommand{\brX}{{\mathbf X}}
\newcommand{\brY}{{\mathbf Y}}
\newcommand{\brZ}{{\mathbf Z}}

\newcommand{\whatH}{\widehat{\mathbf H}}

\newcommand{\whatHk}{\widehat{\mathbf H}_k}
\newcommand{\wtildeHk}{\widetilde{\mathbf H}_k}

\newcommand{\Hk}{{\mathbf H}_k}

\newcommand{\Pk}{{\mathbf P}_k}
\newcommand{\Pj}{{\mathbf P}_j}
\newcommand{\barPk}{{\overline{\mathbf P}}_k}

\newcommand{\barP}{{\overline{\mathbf P}}}

\newcommand{\In}{{\mathbf I}_N}
\newcommand{\sinrk}{\text{SINR}_k}

\newcommand{\Lambdak}{\mathbf{\Lambda}_k}

\newcommand{\No}{\sigma_n^2}
\newcommand{\bbE}{\mathbb E}

\newcommand{\Mkmmse}{\mathbf{M}_k^\text{MMSE}}

\newcommand{\Tr}{\text{tr}\slp}

\newcommand{\RD}{\mathbb{R}}
\newcommand{\CD}{\mathbb{C}}

\definecolor{mydarkblue}{RGB}{37, 24, 181}

\newcommand{\llp}{\left\{}
\newcommand{\lrp}{\right\}}
\newcommand{\slp}{\left(}
\newcommand{\srp}{\right)}
\newcommand{\sll}{\left|}
\newcommand{\srl}{\right|}

\newtheorem{theorem}{Theorem}
\newtheorem{lemma}{Lemma}

\newtheorem{remark}{Remark}

\begin{document}
\title{Robust Precoding Designs for Multiuser MIMO Systems with Limited Feedback}

\author{Wentao~Zhou,~\IEEEmembership{Student Member,~IEEE,}
Di Zhang,~\IEEEmembership{Senior Member,~IEEE,}
Mérouane Debbah,~\IEEEmembership{Fellow,~IEEE,}
and\\ Inkyu Lee,~\IEEEmembership{Fellow,~IEEE}
\thanks{This work was supported by the National Research Foundation of Korea (NRF) funded by the Korea Government (MSIT) under Grants 2022R1A5A1027646. This article was submitted in part at the IEEE VTC 2024 spring \cite{VTC2024}. \textit{(Corresponding author: Inkyu Lee.)}}
\thanks{Wentao Zhou and Inkyu Lee are with the School of Electrical Engineering, Korea University, Seoul 02841, South Korea (e-mail: wtzhou@korea.ac.kr; inkyu@korea.ac.kr).}
\thanks{Di Zhang is with the School of Electrical and Information Engineering, Zhengzhou University, Zhengzhou 450001, China (e-mail: dr.di.zhang@ieee.org).}
\thanks{M\'{e}rouane Debbah is with the Department of Electrical Engineering and Computer Science, Khalifa University of Science and Technology, Abu Dhabi, United Arab Emirates (e-mail: merouane.debbah@ku.ac.ae).}
\thanks{This work has been accepted for publication in \textit{IEEE Transactions on Wireless Communications}.
Personal use of this material is permitted.
Permission from IEEE must be obtained for all other uses, in any current or future media, including reprinting/republishing this material for advertising or promotional purposes, creating new collective works, for resale or redistribution to servers or lists, or reuse of any copyrighted component of this work in other works.}
\thanks{Digital Object Identifier 10.1109/TWC.2024.3363766}
}

\maketitle

\begin{abstract}
It has been well known that the achievable rate of multiuser multiple-input multiple-output systems with limited feedback is severely degraded by quantization errors when the number of feedback bits is not sufficient.
To overcome such a rate degradation, we propose new robust precoding designs which can compensate for the quantization errors.
In this paper, we first analyze the achievable rate of traditional precoding designs for limited feedback systems.
Then, we obtain an approximation of the second-order statistics of quantized channel state information.
With the aid of the derived approximation, we propose robust precoding designs in terms of the mean square error (MSE) with conditional expectation in non-iterative and iterative fashions.
For the non-iterative precoding design, we study a robust minimum MSE (MMSE) precoding algorithm by extending a new channel decomposition.
Also, in the case of iterative precoding, we investigate a robust weighted MMSE (WMMSE) precoding to further improve the achievable rate.
Simulation results show that the proposed precoding schemes yield significant improvements over traditional precoding designs.
\end{abstract}


\IEEEpeerreviewmaketitle

\section{Introduction}
\IEEEPARstart{A} multiuser multiple-input multiple-output (MU-MIMO) system which simultaneously supports several users through the same time and frequency resource is a key technique to improve the achievable rate in wireless communications.
In order to serve different users, a base station (BS) utilizes channel state information (CSI) at the transmitter (CSIT) to conduct the preprocessing process \cite{OVERVIEW08, LFHP15}.
Since the CSIT is not accurate in practice, a degradation of the achievable rate is inevitable if the BS does not compensate for the imperfect CSIT.

In most works, the available CSIT and its associated errors were assumed to be uncorrelated \cite{RSLNR16, RSMISO16, RBDEH19, GPIP20, RBD21, RLTD21, RSMAMO21, RJPD23, BOHI21, MIMIHI23}, where robust techniques were developed by utilizing prior knowledge of an error distribution.
However, the limited feedback system that conveys the CSI back to the BS has a correlated error.
Specifically, the subspace of quantization errors lies in the left nullspace of the available CSIT \cite{LFBD08}.
Thus, correlated channel imperfections caused by the quantization errors should be compensated.

In limited feedback systems, each user quantizes its CSI via a predefined codebook and feeds the chosen index back to the BS.
In the case of multiple users, a mismatch between the quantized CSIT and the channel realization leads to a loss of the achievable rate at high signal-to-noise ratio (SNR) due to interference.
To overcome the rate degradation, several methods have been studied \cite{LFBD08, OFRIL16, ACDC08, CQLF14, SQBC13, MEAR15, OAMG16, OARUS17, OOCSI19, PFIA20, NOMAQO20}.
The quantization error can be made small by simply increasing the number of feedback bits \cite{LFBD08, OAMG16, OFRIL16, OARUS17, OOCSI19}.
However, it may be impractical as the search complexity grows with feedback bits.
Also, the antenna combining techniques in \cite{ACDC08, SQBC13, MEAR15} and new quantization methods in \cite{OAMG16, CQLF14, PFIA20, NOMAQO20, DLCSI23} can reduce the quantization error, while they still fail to further improve the achievable rate.
Thus, it is important to design robust precoding that addresses a degradation of the achievable rate caused by the limited feedback.

Based on conventional algorithms \cite{RZF05, SLNR07, WMMSE08}, some robust precoding schemes have been studied to deal with channel imperfections.
Robust regularized zero-forcing (RZF) precoding designs were examined in \cite{RJPD23, RLTD21, MIMIHI23}, where the regularization terms are determined by extending a conditional expectation of the mean square error (MSE).
By incorporating the second-order statistic of channel realizations, the authors in \cite{RSLNR16, RSLNR18} proposed robust signal-to-leakage-and-noise ratio (SLNR) precoding designs against channel estimation errors.
However, they failed to further improve the system performance since they do not aim to maximize the achievable rate.
A robust weighted minimum mean square error (WMMSE) precoding scheme was presented which iteratively minimizes the MSE to maximize the achievable rate \cite{RWMMSE13, RSMISO16, RSMA22}.
For large transmit antenna systems \cite{PACE13, GLSE19}, robust hybrid precoding designs were examined in \cite{RHP21, RHT23}.
Besides, robust precoding designs based on different optimization frameworks were introduced in \cite{GPIP20, RBDEH19, RBD21, BOHI21}.
However, as these methods are based on uncorrelated channel models, their extension to MU-MIMO systems with limited feedback is nontrivial.
To handle the quantization error, an MMSE precoding was proposed in \cite{RMMSE08} based on the rate distortion theory, where the estimation error variance is replaced by the quantization error variance.
In \cite{RMMSE09} and \cite{EEOLF23}, the correlation between the quantized CSI and its quantization error was utilized to conduct robust precoding designs.
Nevertheless, the above methods work only for multiuser multiple-input single-output (MISO) systems, and cannot be employed in MU-MIMO scenarios.
Actually, the correlated quantization error model was extensively studied in \cite{SQBC13, MEAR15, OAMG16, LRFB16, OARUS17, OOCSI19, QALF21}.
However, only the block-diagonalization (BD) precoding was investigated due to the difficulty of signal processing.
As a result, a precoding design for MU-MIMO systems with limited feedback was left as an open issue \cite{JAPLD22}.

In this paper, we consider new robust precoding schemes for a downlink MU-MIMO system with limited feedback which effectively deals with quantization errors.
The main contributions of this paper are summarized as follows:
\begin{itemize}
\item First, we analyze the achievable rate of the limited feedback system.
In order to overcome a degradation of the achievable rate, we present a tight approximation of the second-order statistics.
We confirm that the derived approximation is quite accurate even for a small number of feedback bits.
Besides, with the approximation in hand, we develop a new channel decomposition model.
\item Based on the proposed approximation, we present robust precoding designs both in non-iterative and iterative fashions.
In the non-iterative case, the conditional expectation based MSE is minimized to conduct the robust MMSE precoding.
Also, for the iterative solution, we solve a weighted sum-rate (WSR) maximization problem by utilizing the weighted MMSE (WMMSE) precoding.
\end{itemize}
The proposed precoding schemes are evaluated by numerical simulations and compared with existing schemes.
The results demonstrate that the proposed methods outperform conventional schemes in the presence of finite feedback bits.

The rest of this paper is organized as follows:
Section \ref{SystemModel} presents the system model and the preliminaries.
The achievable rate of precoding designs and a tight approximation for the second-order statistics are given in Section \ref{ProblemFormulation}.
In Section \ref{RobustPrecoding}, we propose robust precoding designs in non-iterative and iterative fashions.
Besides, the computational complexity is discussed.
Numerical results are exhibited in Section \ref{SimuResults}, and the paper is finally concluded in Section \ref{Conclusion}.

Notation: Lower case bold letters and upper case bold letters represent vectors and matrices, respectively.
$\slp \cdot \srp^{T}$, $\slp \cdot \srp^{H}$, and $\text{tr}\slp \cdot \srp$ stand for transpose, Hermitian, and trace, respectively.
$\text{diag}\llp \brq \lrp$ is a diagonal matrix whose diagonal elements consist of the elements of $\brq$, and the inverse operation $\text{diag}^{-1}\llp \brQ \lrp$ collects all diagonal elements of a matrix $\brQ$ into a column vector.
$\text{blkdiag}\llp \cdot \lrp$ refers to a block diagonal matrix.
$\text{eig}_n \slp \brQ \srp$ indicates the $n$-th eigenvalue of a matrix $\brQ$.
An $N \times N$ identity matrix is denoted by $\mathbf{I}_N$, while $\mathbf{1}_{M \times N}$ refers to an $M \times N$ all-ones matrix.
In addition, $\mathbb{E}\llp \cdot \lrp$ defines the statistical expectation, and $\| \cdot \|_2$ and $\| \cdot \|_F$ denotes the Euclidean and Frobenius norm operations, respectively.

\section{System Model and Preliminaries}
\label{SystemModel}
We consider MU-MIMO systems where a BS equipped with $M$ transmit antennas serves $K$ users each with $N$ receive antennas. 
Denoting $\brH_k \in \CD^{M \times N}$ as the flat-fading channel between the BS and the $k$-th user $\slp k=1,\cdots,K \srp $, the received signal at user $k$ is described as
\begin{equation*}
\begin{aligned}
\bry _k = \brH^H_k \brx + \brn_k,
\end{aligned}
\end{equation*}
where $\brH_k$ consists of independent and identically distributed (i.i.d.) unit variance complex Gaussian random variables, $\brx \in \CD^{M \times 1}$ represents the transmitted signal, and $\brn_k \in \CD^{N \times 1}$ is a zero-mean complex Gaussian noise vector with independently distributed entries of variance $\sigma_n^2$.
The transmitted signal is precoded by $\brx=\sum_{j=1}^{K}\brP_j \brs_j$ where $\brP_k \in \CD^{M \times N}$ and $\brs_k\in\CD^{N \times 1}$ indicate the precoding matrix and the symbol for user $k$, respectively.
To simplify the presentation, we set $M=N K$.
For the downlink broadcast system with limited feedback, it is assumed that channel estimation is perfect, and thus the channel imperfections come only from the quantization processes.

\subsection{Limited Feedback Model}
We adopt the limited feedback model in \cite{LFBD08}, where each user quantizes the subspace of its channel realization via a predefined codebook and feeds $B$ bits back to the BS.
Thus, based on the codebook $\mathcal{C}_k=\{\brC_{k,1}, \cdots, \brC_{k,2^B} \}$ where $\brC_{k,i} \in \CD^{M \times N}$ is the $i$-th codeword in the $k$-th codebook, user $k$ returns index $i$ if $\brC_{k,i}$ is chosen.
Each codeword in $\mathcal{C}_k$ is assumed to be semi-unitary (i.e., $\brC_{k,i}^H \brC_{k,i}= \brI_N$).
Let us denote $\wtildeHk$ as the subspace of $\brH_k$, which can be obtained by eigenvalue decomposition (EVD) of $\brH_k\brH_k^H=\wtildeHk \Lambdak \wtildeHk^H$, where $\Lambdak = \text{diag}\llp \bm{\lambda} \lrp$ indicates the concatenation of unordered eigenvalues with $\bm{\lambda} = \left[\lambda_1,\cdots,\lambda_N\right]^T$.
The quantization is achieved by finding the minimum chordal distance between $\brH_k$ and $\brC_{k,i}$ as \cite{LFBD08}
\begin{equation*}
\begin{aligned}
\whatHk = \argmin_{\brC_{k,i} \in \mathcal{C}_k} d^2 \slp \brH_k,\brC_{k,i} \srp,
\end{aligned}
\end{equation*}
where $d^2 \slp \brH_k,\brC_{k,i} \srp \triangleq N-\text{tr}\slp \wtildeHk^H \brC_{k,i} \brC_{k,i}^H \wtildeHk \srp$.
Thus, the BS can access CSI of all $K$ users after receiving the feedback indices.

Without loss of generality, we adopt a random vector quantization (RVQ) codebook, which serves as a lower bound among all codebook designs.
Although it is possible to obtain better performance from other well-designed codebooks, the RVQ codebook is commonly employed for the sake of analysis.
For a given codebook, the total quantization distortion $\xi$ is defined as
\begin{equation}
\label{RVQe}
\begin{aligned}
\xi = \mathbb{E}\llp \min_{\brC_{k,i} \in \mathcal{C}_k} d^2(\brH_k,\brC_{k,i})  \lrp.
\end{aligned}
\end{equation}
When the number of feedback bits is large, $\xi$ in RVQ can be approximated by
\begin{equation*}
\begin{aligned}
\overline{\xi} = \frac{\Gamma\slp \frac{1}{T}\srp}{T}  C ^{-\frac{1}{T}}2^{-\frac{B}{T}},
\end{aligned}
\end{equation*}
where $\Gamma\slp\cdot\srp$ represents the gamma function, and $T$ and $C$ are defined as $T=N^2\slp K-1 \srp$ and $C=\frac{1}{T!} \prod_{i=1}^N \frac{\slp M-i \srp!}{\slp N-i \srp!}$, respectively.

\subsection{Channel Decomposition}
\label{SecChanDecomp}
The channel decomposition model in \cite{LFBD08} has been extensively studied in \cite{SQBC13, MEAR15, OAMG16, LRFB16, OARUS17, OOCSI19, QALF21}.
According to the quantization process in the rate distortion theory, the essential idea of this decomposition is the projection process of the subspace $\wtildeHk$ onto the column space and the left nullspace of the selected codeword $\whatHk$ as
\begin{equation}
\label{chandecompo}
\begin{aligned}
\wtildeHk =  \whatHk\whatHk^H \wtildeHk + \slp \brI_M - \whatHk\whatHk^H \srp \wtildeHk.
\end{aligned}
\end{equation}
Taking QR decomposition to each term in \eqref{chandecompo} gives
\begin{equation}
\label{chandecomp}
\begin{aligned}
\wtildeHk =  \whatHk \brX_k \brY_k + \brS_k \brZ_k,
\end{aligned}
\end{equation}
where $\brX_k \in \CD^{N \times N}$ represents a unitary matrix, $\brY_k \in \CD^{N \times N}$ and $\brZ_k \in \CD^{N \times N}$ are upper triangular matrices with positive diagonal elements, and an orthonormal basis concatenation $\brS_k \in \CD^{M \times N}$ is isotropically distributed $N$ dimensional plane in the $M-N$ dimensional left nullspace of $\whatHk$.
In addition, $\whatHk$, $\brX_k$, and $\brY_k$ are independent of each other, and so are the pair of $\brS_k$ and $\brZ_k$.
$\brY_k$ and $\brZ_k$ satisfy $\text{tr}\slp \brZ_k^H\brZ_k  \srp=d^2 \slp \brH_k, \whatHk \srp$ and $\brY_k^H\brY_k + \brZ_k^H\brZ_k= \brI_N$.
Denoting the quantization distortion for each column of $\Hk$ as $\gamma = \frac{\xi}{N}$, we have $\bbE\llp \brZ_k^H \brZ_k \lrp = \gamma \In$.

Since the BS requires the instantaneous knowledge of $\brX_k$, $\brY_k$, $\brS_k$, and $\brZ_k$, which are not fed back, it is not a trivial task to design a precoding scheme to mitigate the interference.
It is thus essential to develop robust techniques to improve the achievable rate.
In the following section, we first calculate an approximated achievable rate of the MMSE precoding to demonstrate the limitation of the achievable rate caused by quantization errors.
Then, we propose a tight approximation of the second-order statistics which will be utilized to derive robust precoding designs.

\section{Problem Formulation}
\label{ProblemFormulation}
In MU-MIMO systems, the achievable rate for user $k$ is computed by $R_k = \bbE \llp \log_2\ \sll \In + \sinrk  \srl \lrp$ where the $k$-th signal-to-interference-plus-noise ratio (SINR) is given as
\begin{equation*}
\begin{aligned}
\sinrk = \Hk^H \Pk \Pk^H \Hk \slp \sum_{j \neq k}^{K}\Hk^H \Pj \Pj^H \Hk + \No \In\srp^{-1},
\end{aligned}
\end{equation*}
and the sum rate equals $\sum_k R_k$.
From \eqref{chandecomp}, it is clear that if the precoding for user $k$ is based on the quantized CSI only, quantization errors (i.e., $\brS_k$) yield a degraded achievable rate.
In what follows, we examine the limitation of achievable rate when we design the precoding in terms of the quantized CSI.

Now, the problem for minimizing the MSE is formulated as
\begin{subequations}
\label{MMSEof}
\begin{align}
\label{MMSEofa}
& \underset{\brP}{\text{min}} \
\mathcal{Q}_\text{MMSE}\llp \brP \lrp \triangleq \bbE \llp \| \brs - \beta^{-1}\bry \|_2^2  \lrp \\
\label{MMSEofb}
& \text{subject to} \
\text{tr} \slp \brP \brP^H \srp = \rho,
\end{align}
\end{subequations}
where $\brs = \llp \brs_1^T, \cdots,\brs_K^T \lrp^T$ and $\bry = \llp \bry_1^T, \cdots,\bry_K^T \lrp^T$ represent the transmit symbol vector and the received symbol vector, respectively, $\brP=[\brP_1, \cdots ,\brP_K]$ denotes the concatenation of precoding matrices, and $\beta \in \RD_+$ is a scaling factor satisfying the power constraint.
If perfect CSIT is available, the optimal solution $\Pk$ is obtained from the subspace of $\check{\brP}_k^\text{MMSE}$ in $\check{\brP}^\text{MMSE} = \left[\check{\brP}_1^\text{MMSE},\cdots,\check{\brP}_K^\text{MMSE} \right]$, where $\check{\brP}^\text{MMSE}$ is calculated by \cite{MMSE05}
\begin{equation}
\label{MMSEunnormalize}
\begin{aligned}
\check{\brP}^\text{MMSE} = \slp \brH \brH^H + \frac{M \No}{\rho} \brI_M \srp^{-1} \brH
\end{aligned}
\end{equation}
with $\brH = [\brH_1,\cdots,\brH_k]$.
It is worth noting that we assume equal power allocation since channel quality information (CQI) is not available.
Denoting $\rho$ as the total transmit power, the precoding matrix for user $k$ is normalized as
\begin{equation}
\label{Pknormal}
\begin{aligned}
\Pk = \sqrt{\frac{\rho}{M}} \barPk,
\end{aligned}
\end{equation}
where $\barPk$ represents the subspace of $\check{\brP}_k^\text{MMSE}$.

It can be easily checked that when the transmit power is low, $\brH\brH^H$ in \eqref{MMSEunnormalize} is negligible compared to $\frac{M \No}{\rho} \brI_M$, and thus the MMSE precoding becomes close to maximum ratio transmission (MRT) precoding at low SNR region.
On the contrary, if the transmit power is high, the MMSE precoding approaches the block diagonalization (BD) precoding at high SNR region \cite{LSNLP12}.
However, since accurate channel realizations are not available at the BS, the MMSE precoding cannot compensate for quantization errors.
To analyze the achievable rate in MU-MIMO systems with limited feedback, we introduce the following lemma.

\begin{lemma}
\label{lemma_ergodicequiv}
For positive definite matrices $\brX$, $\brY_1$, $\cdots$, $\brY_J$, we have an approximation as
\begin{equation*}
\begin{aligned}
\bbE\llp \log_2 \sll \brI + \brX \brY^{-1} \srl \lrp \approx \log_2\sll \brI + \bbE \llp \brX \lrp\bbE \llp \brY \lrp^{-1}\srl,
\end{aligned}
\end{equation*}
where $\brY = \sum_{j=1}^J \brY_j$.
The above approximation becomes tight when $J$ increases.
\end{lemma}
\begin{IEEEproof}
Since the proof is similar to the proof of Theorem 1 in \cite{PSUP14} by employing the matrix Taylor series expansion \cite{MatAna16}, we omit the proof for brevity.
\end{IEEEproof}

With Lemma \ref{lemma_ergodicequiv} in hand, we derive approximations for the achievable rate of the MMSE precoding in the following theorem.
\begin{theorem}
\label{theorem_achierate}
In MU-MIMO systems with limited feedback, the achievable rate of each user with the MMSE precoding at the low SNR region is approximated by
\begin{equation*}
\begin{aligned}
R^\text{low} = N \log_2 \slp 1+ \frac{\rho K \slp 1 - \gamma \srp}{\rho K - \rho + K} \srp.
\end{aligned}
\end{equation*}
Besides, at the high SNR region, the achievable rate of each user becomes
\begin{equation*}
\begin{aligned}
R^\text{high}  =  N \log_2 \slp 1+ \frac{\rho}{K + \rho K \gamma} \srp.
\end{aligned}
\end{equation*}
\end{theorem}
The above approximations get tight as the number of transmit antennas or users increases.
\begin{IEEEproof}
See Appendix \ref{appendix_theorem_achierate}.
\end{IEEEproof}

\begin{figure}[t]
\centering
\includegraphics[width=3.5in]{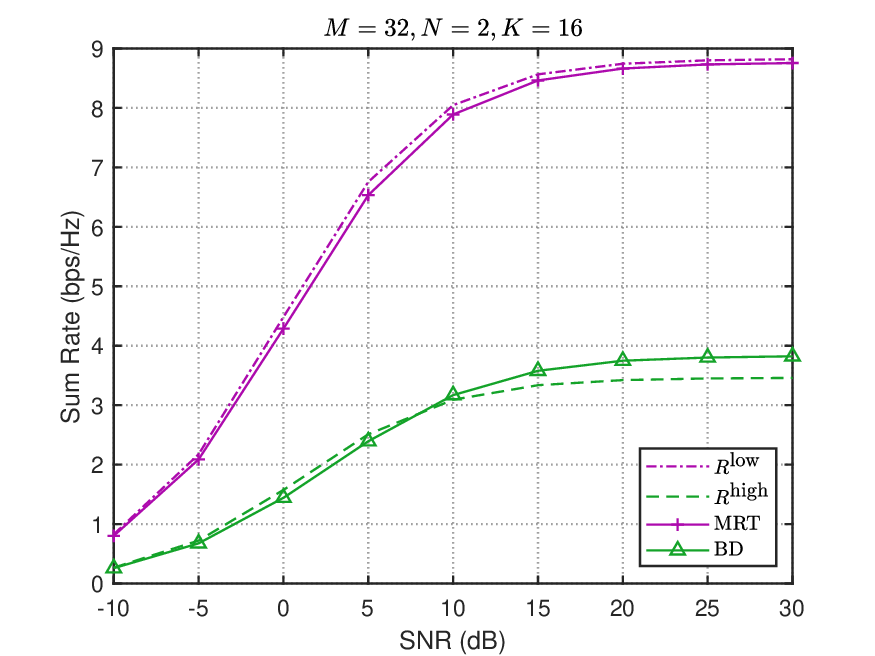}
\caption{Achievable rate of the MRT precoding and the BD precoding with their bounds under $B=10$.}
\label{fig_boundcomp}
\end{figure}

The comparison between the proposed bounds and the MRT and BD precoding schemes is illustrated in Fig. \ref{fig_boundcomp}.
In Theorem \ref{theorem_achierate}, we can see that the achievable rate is saturated independent of the transmit power at high SNR region, i.e., $\lim_{\text{SNR}\rightarrow \infty} R^\text{high} = N \log_2 \slp 1+ \frac{1}{K\gamma} \srp$.
To overcome this rate degradation, the number of feedback bits needs to be increased to scale down the quantization distortion $\gamma$ \cite{LFBD08}.
However, this may not be a feasible solution as the search complexity grows with the number of feedback bits.

Now, we examine the second-order statistic $\brR_k=\bbE\llp \Hk\Hk^H \lrp$.
First, we take EVD to $\Hk \Hk^H = \wtildeHk \Lambdak \wtildeHk^H$.
Recalling the channel decomposition model \eqref{chandecomp}, the second-order statistic is decomposed as
\begin{equation}
\label{Rk}
\begin{aligned}
\brR_k = & \bbE \llp \slp \whatHk \brX_k \brY_k + \brS_k \brZ_k \srp \Lambdak \slp \whatHk \brX_k \brY_k + \brS_k \brZ_k \srp^H \lrp \\
= & M \bbE \llp \whatHk \brX_k \brY_k \brY_k^H \brX_k^H \whatHk^H \lrp + M \bbE \llp \brS_k \brZ_k \brZ_k^H \brS_k^H \lrp \\
& + \underbrace{M \bbE \llp \brS_k \brZ_k \brY_k^H \brX_k^H \whatHk^H \lrp}_{\brR_{k,\brZ\brY}} + \underbrace{M \bbE \llp \whatHk \brX_k \brY_k \brZ_k^H \brS_k^H \lrp}_{\brR_{k,\brY\brZ}},
\end{aligned}
\end{equation}
where the last equality follows from the fact that $\Lambdak$ is independent of other components and $\bbE \llp \Lambdak \lrp = M \brI_N$.
In general, if CSI and its error are independent of each other, $\brR_k$ can be determined simply by summing up their distributions.
However, we cannot utilize this to derive $\brR_k$ since $\brY_k$ and $\brZ_k$ are correlated.
Thus, we investigate an approximation to simplify $\brR_k$ based on the following lemma.

\begin{lemma}
\label{lemma_equality}
For a random unitary matrix $\brX \in \CD^{N \times N}$ and a random matrix $\brY \in \CD^{N \times N}$ with i.i.d. entries and $\bbE \llp \brY \brY^H \lrp = \text{diag}\llp \bm{\lambda}^\bry \lrp$ where $\bm{\lambda}^\bry = \left[\lambda_{1}^\bry,\cdots,\lambda_{N}^\bry \right]^T$, we have the following equality if $\brX$ and $\brY$ are independent of each other as
\begin{equation}
\label{lemma_eq}
\begin{aligned}
\bbE \llp \brY^H \brX^H \brX \brY \lrp = \bbE \llp \brX \brY \brY^H \brX^H \lrp.
\end{aligned}
\end{equation}
\end{lemma}
\begin{IEEEproof}
See Appendix \ref{appendix_lemma_equality}.
\end{IEEEproof}
Based on Lemma \ref{lemma_equality}, we obtain an approximation for $\brR_k$ in the following.

\begin{theorem}
\label{theorem_Rk}
For a small quantization distortion $\gamma$ or a large number of transmit antennas, $\brR_k$ for limited feedback MU-MIMO systems with semi-unitary codebooks can be approximated by
\begin{equation}
\label{chancovmtx}
\begin{aligned}
\brR_{k}^o = M \slp 1-\frac{M\gamma}{M-N} \srp \whatHk \whatHk^H + \frac{MN\gamma}{M-N} \brI_M.
\end{aligned}
\end{equation}
\end{theorem}
\begin{IEEEproof}
See Appendix \ref{appendix_theorem_Rk}.
\end{IEEEproof}

\begin{figure}[t]
\centering
\includegraphics[width=3.5in]{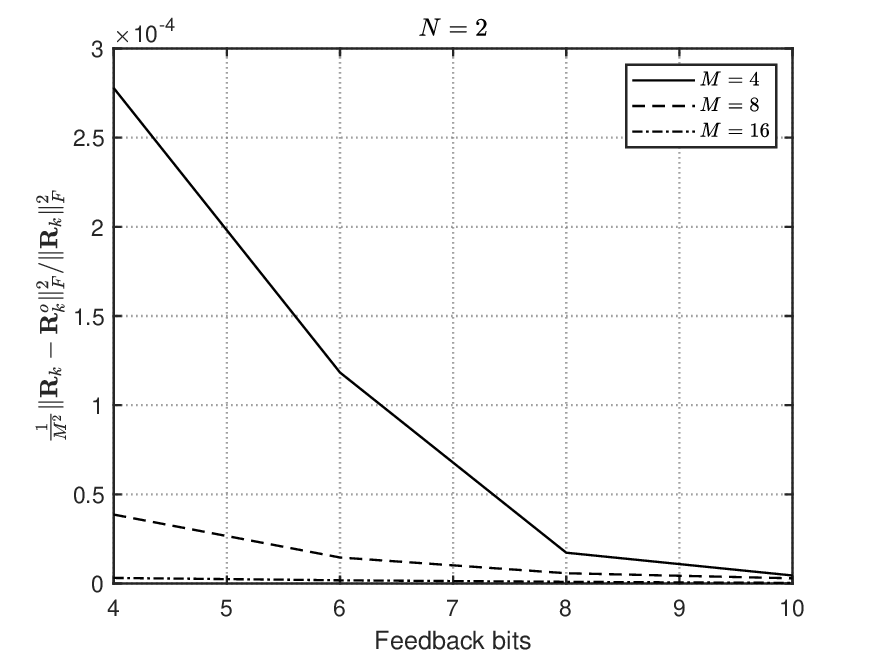}
\caption{Normalized gap between $\brR_k$ and $\brR_k^o$ with respect to the number of feedback bits.}
\label{fig_DiffBwithFixedM_RkandRko}
\end{figure}

The normalized gap between $\brR_k$ and $\brR_k^o$ is illustrated in Fig. \ref{fig_DiffBwithFixedM_RkandRko}.
It is clear that the difference between $\brR_k$ and $\brR_k^o$ becomes negligible as $B$ and $M$ increase.
One should also notice that although a small quantization distortion assumption is adopted in Theorem \ref{theorem_Rk}, the approximation \eqref{chancovmtx} is quite accurate even for a small $B$.
Note that an approximation $\gamma \approx \overline{\xi}/N $ holds for the RVQ codebook.
With the derived approximation, we introduce robust precoding designs to compensate for quantization errors in the next section.

\section{Robust Precoding Designs}
\label{RobustPrecoding}
In this section, we propose new robust precoding designs in both non-iterative and iterative fashions.
First, for the non-iterative precoding, the conditional expectation based MSE is minimized to compensate for quantization errors.
Next, we alternatively minimize the weighted MSE with conditional expectation in the iterative solutions.
We will show that the iterative precoding design further improves the achievable rate compared to the non-iterative precoding scheme.

\subsection{Robust Non-iterative Precoding Design}
The conventional MMSE precoding design involves taking the expectation on the transmitted signals and noise.
However, since some matrices in the channel decomposition model \eqref{chandecomp} are not known, the conventional MMSE precoding design becomes infeasible in limited feedback systems.
To overcome this problem, one may take the conditional expectation over known CSI and its error.
In this paper, the optimization problem for robust MMSE (RMMSE) can be obtained by taking the conditional expectation over the quantized CSI as
\begin{subequations}
\label{MMSEeq}
\begin{align}
\label{MMSEeqa}
& \underset{\brP}{\text{min}} \
\bbE_{\brH|\whatH} \llp \mathcal{Q_\text{MMSE}\llp \brP \lrp} \lrp \\
\label{MMSEeqb}
& \text{subject to}\
\text{tr} \slp \brP \brP^H \srp = \rho.
\end{align}
\end{subequations}
To determine the optimal $\brP$, we apply the Lagrange multiplier method as \cite{CO04}
\begin{equation}
\begin{aligned}
L \slp\brP, \beta,\lambda \srp \! = & \bbE_{\brH|\whatH} \llp \mathcal{Q}_\text{MMSE}\llp \brP \lrp  \lrp + \lambda \slp \text{tr} \slp \brP \brP^H \srp - \rho\srp \\
 = & M \! - \! \frac{1}{\beta} \Tr \! \brP^H \! \bbE _{\brH|\whatH}\! \llp \brH \lrp \! \srp \! - \! \frac{1}{\beta} \Tr \! \brP \bbE_{\brH|\whatH}\! \llp \brH^H \lrp \! \srp \\
 & + \frac{1}{\beta^2} \Tr \brR \brP \brP^H \srp + \frac{M \No}{\beta^2} + \lambda \Tr \brP \brP^H \srp  - \lambda \rho ,
\end{aligned}
\end{equation}
where $\brR=\sum_{j=1}^K \bbE _{\brH|\whatH} \llp \brH_j \brH_j^H \lrp$ represents the concatenation of channel covariances and $\lambda \in \RD_+$ is the Lagrange multiplier.
The optimal solution can be obtained by solving $\frac{\partial L \slp\brP, \beta,\lambda \srp}{\partial \brP^H}=\mathbf{0}$ and $\frac{\partial L \slp\brP, \beta,\lambda \srp}{\partial \beta}=\mathbf{0}$ as
\begin{equation}
\label{MMSEde}
\begin{aligned}
\check{\brP}^\text{RMMSE} = \beta \slp \brR +\frac{M\No}{\rho} \brI_M \srp^{-1} \bbE _{\brH|\whatH} \llp\brH \lrp.
\end{aligned}
\end{equation}
However, both $\brR$ and $\bbE_{\brH|\whatH} \llp\brH \lrp$ are unknown to the BS.
In the following, we explore a method to compute these two matrices.
First, we can utilize Theorem \ref{theorem_Rk} to approximate $\brR$ as  $\brR^o = \sum_{j=1}^K \brR_j^o$.
Next, for $\bbE_{\brH|\whatH} \llp\brH \lrp$, we can see that from the channel decomposition model \eqref{chandecomp}, $\bbE _{\brH|\whatH} \llp\brH \lrp$ consists of the concatenation of $\whatHk \brX_k \brY_k$ and $\brS_k \brZ_k$.
Since $\brX_k$, $\brY_k$, $\brS_k$, and $\brZ_k$ are unknown at the BS, we then have the following lemma.

\begin{lemma}
\label{lemma_chandecomp}
For a small quantization distortion $\gamma$ or a large number of transmit antennas, the channel realization $\Hk$ can be approximated by
\begin{equation}
\label{chandecomp2}
\begin{aligned}
\Hk \approx \delta_k \whatHk + \eta_k \brO_k,
\end{aligned}
\end{equation}
where $\delta_k$ and $\eta_k$ indicate random values with $\delta \triangleq \bbE \llp \delta_k \lrp = \sqrt{M -\frac{M^2\gamma}{M-N}}$ and $\eta \triangleq \bbE \llp \eta_k \lrp = \sqrt{\frac{M^2\gamma}{M-N}}$, and $\brO_k$ is a random semi-unitary matrix independent of $\whatHk$.
Each entry in $\brO_k$ is an i.i.d. complex Gaussian random variable with variance of $\frac{1}{M}$.
Besides, the conditional expectation of $\Hk$ can be approximated as
\begin{equation*}
\bbE_{\brH|\whatH} \llp\Hk\lrp \approx \delta \whatHk.
\end{equation*}
\end{lemma}
\begin{IEEEproof}
See Appendix \ref{appendix_lemma_chandecomp}.
\end{IEEEproof}

The proposed channel decomposition model approximates \eqref{chandecomp} by adjusting $\delta$ and $\eta$.
For a large $B$, $\delta$ and $\eta$ approach $\sqrt{M}$ and 0, respectively, which becomes equivalent to the perfect CSI.
Based on Lemma \ref{lemma_chandecomp}, the RMMSE precoding is summarized as
\begin{equation}
\label{robustMMSE}
\begin{aligned}
\check{\brP}^\text{RMMSE} = \slp \brR^o + \frac{M \No}{\rho} \brI_M \srp^{-1} \whatH,
\end{aligned}
\end{equation}
where $\check{\brP}^\text{RMMSE} = \left[\check{\brP}_1^\text{RMMSE},\cdots,\check{\brP}_K^\text{RMMSE} \right]$ and $\whatH=[\whatH_1,\cdots,\whatH_K]$.
Similar to the conventional MMSE precoding scheme, the RMMSE precoding approaches MRT at low SNR region.
In addition, the RMMSE precoding becomes a robust version of the BD precoding at high SNR region.
We will show in the simulation section that a degradation of the achievable rate can be alleviated by the proposed non-iterative precoding.
Next, we alternatively minimize the weighted MSE with conditional expectation to further improve the achievable rate in the following subsection.

\subsection{Robust Iterative Precoding Design}
\label{RWMMSE}
Under perfect CSIT, the WSR  problem is given by
\begin{subequations}
\label{WSR}
\begin{align}
& \underset{\brP}{\text{max}} \
\sum_{k=1}^K \mu_k R_k \\
& \text{subject to} \
\text{tr} \slp \brP \brP^H \srp = \rho,
\end{align}
\end{subequations}
where $\mu_k$ is a predefined weight for user $k$.
Our objective is to find the optimal $\brP$ to maximize the WSR with quantized CSI.
However, solving \eqref{WSR} is not a trivial task, especially when CSI is not perfect.
Since the WSR maximization problem is equivalent to a weighted MSE minimization by introducing a weight matrix $\brW_k$ \cite{WMMSE08}, in this paper, we optimize the WMMSE problem instead.
Let us define $\Mkmmse$ and $\brW_k=\mu_k\slp \Mkmmse \srp^{-1}$ as the minimum MSE matrix and the weight matrix of user $k$, respectively.
The optimization problem of the WMMSE precoding design is formulated as
\begin{subequations}
\label{WMMSEof}
\begin{align}
\label{WMMSEofa}
& \underset{\brP}{\text{min}} \
\sum_{k=1}^K \Tr \brW_k \Mkmmse \srp \\
\label{WMMSEofb}
& \text{subject to} \
\text{tr} \slp \brP \brP^H \srp = \rho.
\end{align}
\end{subequations}
The above problem can be resolved using the block coordinate descent (BCD) method, which iteratively updates $\Mkmmse$ and $\brW_k$ \cite{NP99}.
In what follows, we will derive expressions for the robust WMMSE (RWMMSE) precoding.

Denoting $\brD_k$ as the receive filter at user $k$, the MSE matrix $\brM_k$ between the transmitted data symbol $\brs_k$ and its estimated value $\widehat{\brs}_k=\brD_k \bry_k$ is computed as
\begin{equation}
\label{Mkorig}
\begin{aligned}
\brM_k  = \ & \bbE \llp \slp \widehat{\brs}_k- \brs_k \srp\slp \widehat{\brs}_k-\brs_k \srp^H  \lrp \\
 = \ & \brD_k \Hk^H \brP \brP^H \Hk \brD_k^H - \brD_k \Hk^H \Pk  - \Pk^H \Hk\brD_k^H \\
 & + \No\brD_k\brD_k^H + \brI_N.
\end{aligned}
\end{equation}
To obtain the MMSE matrix $\brM_k^\text{MMSE}$, we must first calculate the receive filter $\brD_k$.
However, it is infeasible to derive $\brD_k$ and $\brM_k^\text{MMSE}$ from \eqref{Mkorig} as $\Hk$ is unavailable in limited feedback systems.
Thus, for a given $\whatHk$, in order to make sure that MSE is minimized for each realization of $\brS_k$, we take the conditional expectation of $\brM_k$ over quantized CSI \cite{RWMMSE13}.
Therefore, we formulate a WMMSE problem with conditional expectation as
\begin{subequations}
\label{WMMSErof}
\begin{align}
\label{WMMSErofa}
& \underset{\brP}{\text{min}} \
\Tr \bbE_{\brH|\whatH} \llp \brW \brM \lrp \srp \\
\label{WMMSErofb}
& \text{subject to} \
\text{tr} \slp \brP \brP^H \srp = \rho,
\end{align}
\end{subequations}
where $\brW = \text{blkdiag} \llp \brW_1,\cdots,\brW_K\lrp$ and $\brM = \text{blkdiag} \llp \brM_1^\text{MMSE},\cdots,\brM_K^\text{MMSE}\lrp$.
Since $\brW$ is a function of $\brM$, we first determine the minimum MSE matrix with conditional expectation, and then compute the new weight matrix.

Recalling the channel decomposition model \eqref{chandecomp2} in Lemma \ref{lemma_chandecomp}, the conditional expectation based MSE matrix $\overline{\brM}_k= \bbE_{\Hk | \whatHk} \llp \brM_k \lrp$ based on \eqref{chandecomp2} is written as
\begin{equation}
\label{ave_mse}
\begin{aligned}
 \overline{\brM}_k = \ & \delta^2 \brD_k \whatHk^H \brP \brP^H \whatHk \brD_k^H + \eta^2\bbE\llp \brD_k \brO_k^H \brP \brP^H \brO_k \brD_k^H \lrp \\
& - \delta \brD_k \whatHk^H \Pk - \delta \Pk^H \whatHk\brD_k^H + \No\brD_k\brD_k^H + \brI_N,
\end{aligned}
\end{equation}
where the expectation is taken with respect to quantization errors.
As we can see, the second item in \eqref{ave_mse} is undetermined because of the random matrix $\brO_k$.
We introduce the following lemma to deal with it.
\begin{lemma}
\label{lemma_expectation}
Let us denote $\brX \in \CD^{M \times M}$ as a fixed matrix and $\brY \in \CD^{M \times N}$ as a random matrix with i.i.d. entries of zero mean and variances $\sigma_{mn}^2, \forall \ m,n$.
Assuming that $\brX$ and $\brY$ are independent, we have
\begin{equation*}
\begin{aligned}
\bbE \llp \brY^H \brX \brY \lrp = \text{diag}\llp \mathbf{\Theta} \text{diag}^{-1}\llp \brX \lrp \lrp,
\end{aligned}
\end{equation*}
where $\mathbf{\Theta} \in \RD^{N \times M}$ is a matrix whose $\slp n,m \srp$-th element equals $\sigma_{mn}^2$.
\end{lemma}
\begin{IEEEproof}
See Appendix \ref{appendix_lemma_expectation}.
\end{IEEEproof}

From Lemma \ref{lemma_expectation}, the expectation of the quantization error in \eqref{ave_mse} can be given as
\begin{equation*}
\begin{aligned}
\bbE\llp \brD_k \brO_k^H \brP \brP^H \brO_k \brD_k^H \lrp = \brD_k  \mathbf{\Phi}_k  \brD_k^H,
\end{aligned}
\end{equation*}
where $\mathbf{\Phi}_k \triangleq \text{diag} \llp \mathbf{\Theta}_k \text{diag}^{-1}\llp \brP\brP^H \lrp \lrp$ and $\mathbf{\Theta}_k = \frac{1}{M} \mathbf{1}_{N \times M}$.
It is worth noting that each element in $\brO_k$ is independent of $\whatHk$.
Therefore, the MSE matrix with conditional expectation can be expressed as
\begin{equation}
\label{ave_mse2}
\begin{aligned}
\overline{\brM}_k  = \ & \delta^2 \brD_k \whatHk^H \brP \brP^H \whatHk \brD_k^H + \eta^2 \brD_k  \mathbf{\Phi}_k  \brD_k^H  - \delta \brD_k \whatHk^H \Pk \\
& - \delta \Pk^H \whatHk\brD_k^H + \No\brD_k\brD_k^H + \brI_N.
\end{aligned}
\end{equation}
Then, the receive filter is obtained by solving $\frac{\partial \overline{\brM}_k}{\partial \brD_k^H}=\mathbf{0}$, which yields
\begin{equation}
\label{rfmmse}
\begin{aligned}
\brD_k^\text{MMSE} = \delta \Pk^H \whatHk \brF_k^{-1},
\end{aligned}
\end{equation}
where $\brF_k=\delta^2 \whatHk^H \brP \brP^H \whatHk +\eta^2 \mathbf{\Phi}_k +\No \brI_N$.
Plugging $\brD_k^\text{MMSE}$ into $\overline{\brM}_k$, the minimum MSE with conditional expectation becomes
\begin{equation}
\label{ave_mse3}
\begin{aligned}
\overline{\brM}_k^\text{MMSE} = \brI_N - \delta^2 \Pk^H \whatHk \brF_k^{-1}\whatHk^H \Pk.
\end{aligned}
\end{equation}
It is essential to distinguish between the receive filter computed in \eqref{rfmmse} and the optimal receive filters.
The former minimizes the conditional MSE based on quantized CSI, while the latter minimizes the actual MSE based on the true channel.
Therefore, the receive filter in \eqref{rfmmse} is derived for the iterative precoding design under imperfect CSI.
This rationale also applies to the minimum MSE matrices based on conditional expectation.

Now, we determine the optimum weight matrix.
In order to ensure the equivalence between the WSR and WMMSE problems, the optimum weight matrix is given by \cite{RWMMSE13}
\begin{equation}
\label{weightcon}
\begin{aligned}
\overline{\brW} = \text{blkdiag} \llp \mu_1\slp \overline{\brM}_1^\text{MMSE} \srp^{-1},\cdots,\mu_K\slp \overline{\brM}_K^\text{MMSE} \srp^{-1}\lrp.
\end{aligned}
\end{equation}
With the MMSE receive filter and the minimum MSE matrix in hand, we rewrite the optimization problem \eqref{WMMSErof} to obtain the precoding matrix $\brP$ as
\begin{subequations}
\label{WMMSErof2}
\begin{align}
\label{WMMSErof2a}
& \underset{\brP}{\text{min}} \
\Tr \overline{\brW} \ \overline{\brM} \srp \\
\label{WMMSErof2b}
& \text{subject to} \
\text{tr} \slp \brP \brP^H \srp = \rho,
\end{align}
\end{subequations}
where $\overline{\brM}$ is defined as
\begin{equation}
\label{MSEmmse}
\begin{aligned}
\overline{\brM} = & \delta^2 \brD \whatH^H \brP \brP^H \whatH \brD^H + \eta^2 \bbE\llp \brD \brO^H \brP \brP^H \brO \brD^H \lrp \\
& - \delta \brD \whatH^H \brP - \delta \brP^H \whatH\brD^H + \No\brD\brD^H + \brI_{KN}
\end{aligned}
\end{equation}
with $\brD=\text{blkdiag}\llp \brD_1^\text{MMSE},\cdots,\brD_K^\text{MMSE} \lrp$ and $\brO=[\brO_1,\cdots,\brO_K ]$.
It is worth noting that as the optimum receive filter $\brD$ can be obtained, $\overline{\brM} $ is actually the minimum conditional expectation based MSE with unknown error subspaces $\brO$ which is a function of $\brP$.
Plugging \eqref{MSEmmse} into $\Tr \overline{\brW} \ \overline{\brM} \srp$, we need to determine $\Tr \overline{\brW} \bbE\llp \brD \brO^H \brP\brP^H \brO\brD^H \lrp\srp$ which can be computed from Lemma \ref{lemma_expectation} as
\begin{equation}
\label{mani2}
\begin{aligned}
\Tr \overline{\brW} \bbE \llp \brD \brO^H \brP \brP^H \brO \brD^H \lrp \srp & \! = \! \Tr \bbE \llp \brO \brD^H \overline{\brW} \brD \brO^H \lrp \brP \brP^H \srp \\
& = \! \Tr \mathbf{\Phi}\brP \brP^H\srp,
\end{aligned}
\end{equation}
where $\mathbf{\Phi}=\text{diag} \llp \mathbf{\Theta} \text{diag}^{-1}\llp\brD^H\overline{\brW}\brD \lrp \lrp$ with $\mathbf{\Theta}=\frac{1}{M}\mathbf{1}_{M \times NK}$.
Finally, we utilize the Lagrange multiplier method to obtain the unnormalized precoding matrix as
\begin{equation}
\label{Pwmmse}
\begin{aligned}
\overline{\brP} = \mathbf{\brT}^{-1}\whatH\brD^H\overline{\brW},
\end{aligned}
\end{equation}
where $\mathbf{\brT} = \delta^2\whatH\brD^H\overline{\brW}\brD\whatH^H+\eta^2 \mathbf{\Phi} +\frac{\No}{\rho}\Tr\overline{\brW}\brD\brD^H\srp\brI_M$.
Then, RWMMSE precoding is obtained by
\begin{equation}
\label{Pwmmsep}
\begin{aligned}
\brP^\text{RWMMSE}= \sqrt{\rho/\Tr \overline{\brP}^{H} \overline{\brP} \srp} \overline{\brP},
\end{aligned}
\end{equation}
where $\brP^\text{RWMMSE} = \left[\brP_1^\text{RWMMSE},\cdots,\brP_K^\text{RWMMSE} \right]$.
To optimize the MMSE receive filter, the weight matrix and the RWMMSE precoding, an iterative algorithm is adopted which alternatively updates each matrix until convergence.
The procedure of the proposed robust iterative precoding is summarized in Algorithm \ref{RobustWMMSE}.

\begin{algorithm}  
\caption{Robust Weighted MMSE Precoding}  
\label{RobustWMMSE}
\begin{algorithmic}
\STATE {Initialize $\brP=\brP^\text{init}$, $\delta = \sqrt{M -\frac{M^2\gamma}{M-N}}$, $\eta = \sqrt{\frac{M^2\gamma}{M-N}}$}
\REPEAT
\STATE {$\brD_k^\text{MMSE} \leftarrow \delta \Pk^H \whatHk \brF_k^{-1}$, where $\brF_k = \delta^2 \whatHk^H \brP \brP^H \whatHk +\eta^2 \mathbf{\Phi}_k +\No \brI_N$, $\forall k$.}
\STATE {$\overline{\brM}_k^\text{MMSE} \leftarrow \brI_N - \delta^2 \Pk^H \whatHk \brF_k^{-1}\whatHk^H \Pk$, $\forall k$}
\STATE {$\overline{\brP} \leftarrow \brT^{-1} \whatH\brD^H\overline{\brW}$, where $\brT = \delta^2\whatH\brD^H\overline{\brW}\brD\whatH^H+\eta^2 \mathbf{\Phi} +\frac{\No}{\rho}\Tr\overline{\brW}\brD\brD^H\srp\brI_M$}
\STATE {$\brP^\text{RWMMSE} \leftarrow \sqrt{\rho/\Tr \overline{\brP}^{H} \overline{\brP} \srp} \overline{\brP} $}
\UNTIL{convergence}
\end{algorithmic}  
\end{algorithm}

\remark
Observing the proposed non-iterative and iterative precoding designs, it is evident that the former is a special case of the latter.
Specifically, if we turn off the iteration, the calculation of precoding matrices \eqref{robustMMSE} and \eqref{Pwmmse} becomes identical.
Both of them are extensions of the RZF precoding which adapts the regularization factor in terms of the system setup and transmit power.
It is worth noting that applying conventional robust precoding schemes in the limited-feedback MU-MIMO system is nontrivial.
This arises from the difficulty of determining the terms associated with the quantization error.
From this perspective, we compare the proposed precoding designs with traditional precoding schemes in the next section.

\subsection{Computational Complexity Analysis}
\label{CompCompAna}
In this subsection, we analyze the computational complexities of the proposed precoding designs in terms of the floating-point operations (flops) \cite{FLOP07}.
The RMMSE precoding design \eqref{robustMMSE} involves basic matrix operations such as multiplication, addition, and inversion.
Then, the power allocation requires the QR decomposition to obtain the subspace of each precoding matrix.
In summary, the proposed RMMSE precoding needs $M^3 + M^2\slp 2NK+0.5K+1 \srp + M\slp 3N^2 K + 3NK -0.5K +3 \srp - \frac{2}{3}K\slp N^3+1 \srp$ flops.

In constrast, for the RWMMSE precoding design, each iteration consists of the following steps:
\begin{itemize}
\item The update of the receive filters \eqref{rfmmse} requests $NK\slp 2M^2+5MN+2M+3N^2-2.5N+1.5 \srp$ flops.
\item Since the conditional expectation based MSE can be rewritten as $\overline{\brM}_k^\text{MMSE} = \brI_N - \delta\brD_k^\text{MMSE}\whatHk^H \Pk$, the update of weight matrices \eqref{weightcon} needs $NK\slp 5N^2-N+2 \srp$ flops.
\item The update of the precoding matrix \eqref{Pwmmsep} is computed in $M^3+4M^2NK+5MN^2K^2+MNK+3M+4N^3K^3-2.5N^2K^2 + 1.5NK-2$ flops.
\end{itemize}

\section{Simulation Results}
\label{SimuResults}
In the numerical simulations, we adopt 5000 channel realizations, where each realization consists of i.i.d. entries with zero mean and unit variance.
Additionally, the noise variance is fixed to 1.
The iterative precoding algorithms are initialized with the MRT precoding and equal weights.
If not specified otherwise, we set the number of feedback bits to 10.
We compare the proposed robust precoding designs with MRT, BD \cite{LFBD08}, MMSE \cite{MMSE05}, and WMMSE \cite{WMMSE08}.
Since each user only feeds the quantized subspace back to the BS, the channel for nonrobust precoding schemes can be given by $\Hk=\sqrt{M}\whatHk$.
In addition, perfect CSI is obtained by $\wtildeHk$.

\begin{figure}[t]
\centering
\includegraphics[width=3.5in]{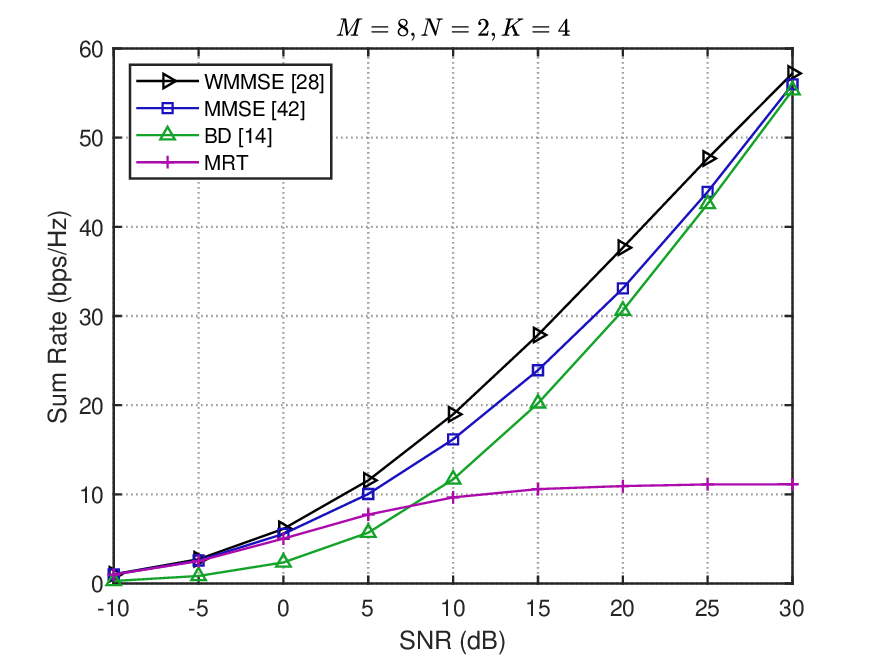}
\caption{Achievable rate of MU-MIMO systems with perfect CSI.}
\label{fig_InfBits}
\end{figure}

First of all, we compare the achievable rate under perfect CSIT in Fig. \ref{fig_InfBits}.
It is worth noting that since the quantization distortion $\gamma$ equals 0, the proposed RMMSE and RWMMSE boil down to MMSE and WMMSE, respectively.
In the plot, BD shows a non-negligible rate degradation over MMSE and WMMSE as BD does not compensate for the noise.
The achievable rate of MMSE displays better performance than that of MRT and BD, and WMMSE exhibits the best performance among all precoding schemes.
We can check that the gap between MMSE and WMMSE is small at low and high SNR regions.
The reason is that due to the lack of CQI, WMMSE cannot allocate power properly, and thus WMMSE degrades into MRT at low SNR region and BD at high SNR region.

\begin{figure}[t]
\centering
\includegraphics[width=3.5in]{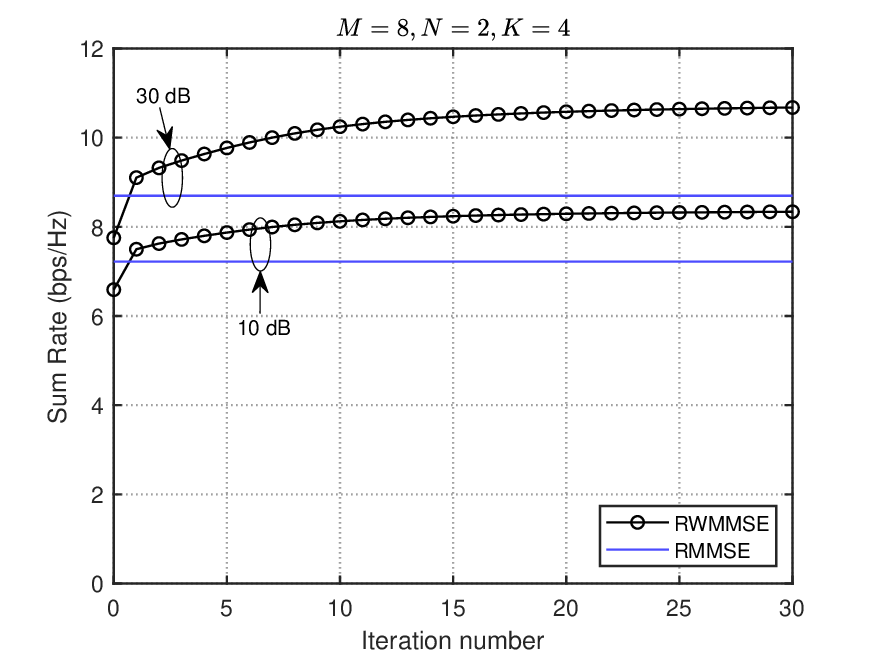}
\caption{Convergence behavior of the proposed iterative precoding design.}
\label{fig_Convergence}
\end{figure}

Next, we investigate the convergence behavior of the proposed iterative precoding design in Fig. \ref{fig_Convergence}. 
From the plot, it can be seen that the RWMMSE precoding design outperforms RMMSE by 21.5 \% for SNR = 30 dB after convergence.
Besides, the RWMMSE precoding design converges within 20 iterations.

\begin{figure}[t]
\centering
\subfigure[$M=8$, $N=2$, $K=4$]{
\label{fig_SR_10bits_824}
\includegraphics[width=3.5in]{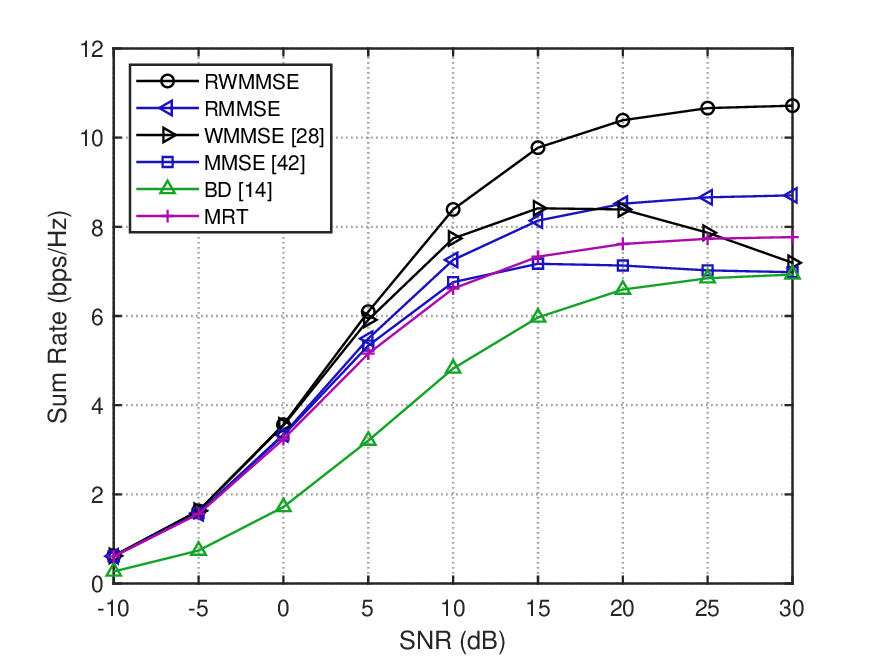}}
\subfigure[$M=16$, $N=2$, $K=8$]{
\label{fig_SR_10bits_1628}
\includegraphics[width=3.5in]{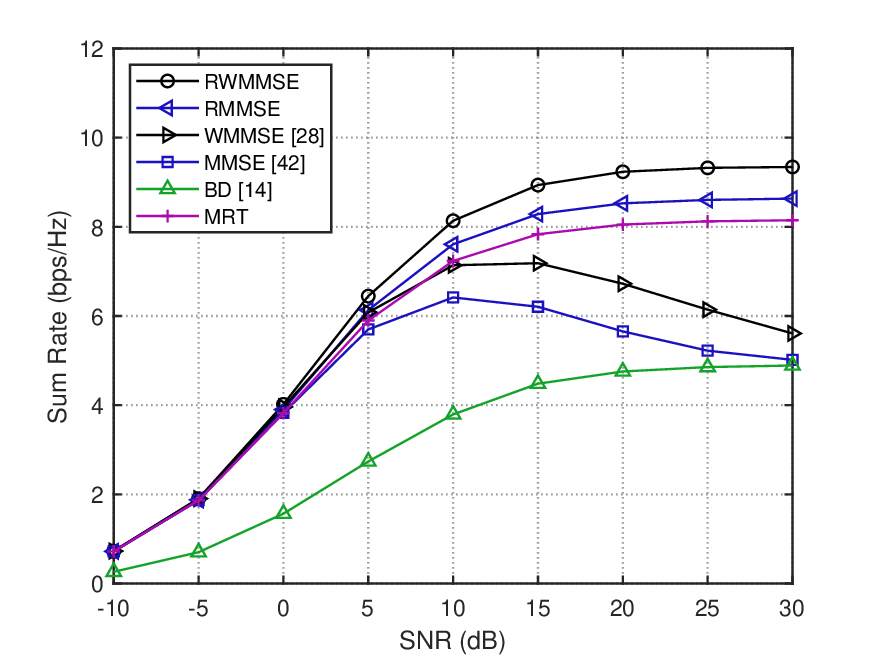}}
\caption{Achievable rate of limited-feedback MU-MIMO systems with $B=10$.}
\label{fig_MMSEandRMMSE}
\end{figure}

\begin{figure}[t]
\centering
\subfigure[SNR = 10 dB]{
\label{fig_SR_VarB_10dB}
\includegraphics[width=3.5in]{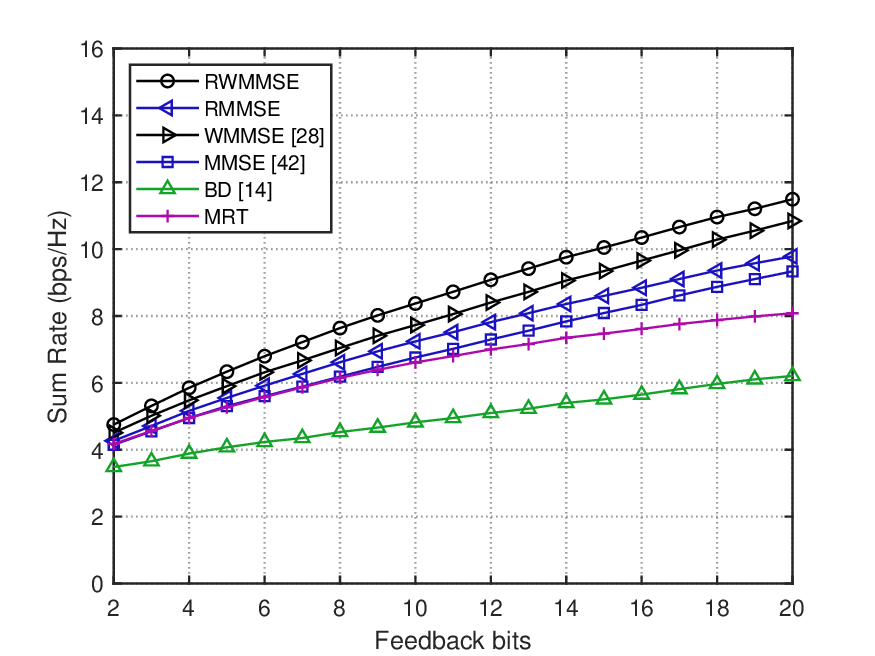}}
\subfigure[SNR = 30 dB]{
\label{fig_SR_VarB_30dB}
\includegraphics[width=3.5in]{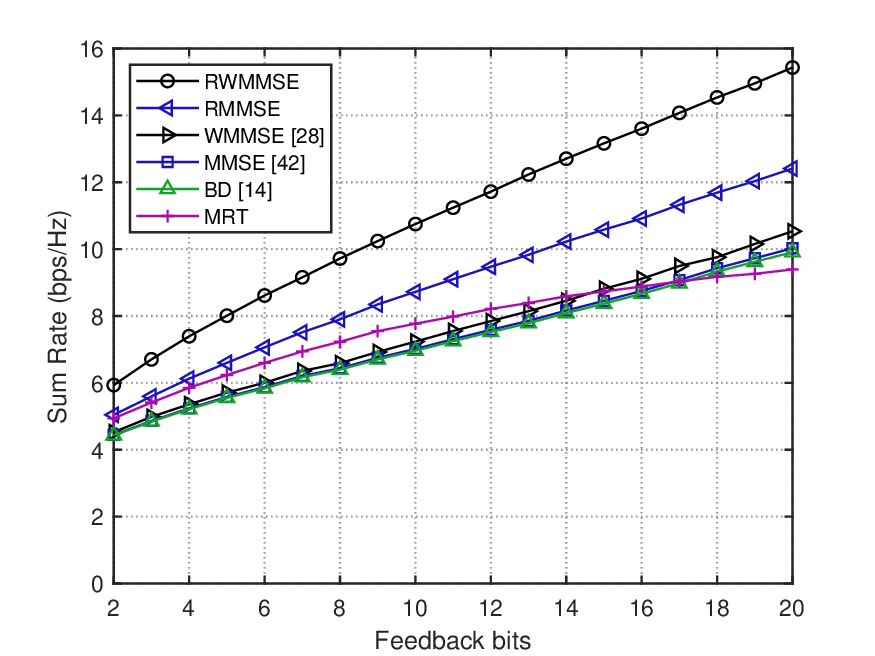}}
\caption{Achievable rate of limited-feedback MU-MIMO systems with $M=8$, $N=2$ and $K=4$.}
\label{fig_SR_VarB}
\end{figure}

We then compare the achievable rate performance of the proposed precoding designs and conventional schemes in Fig. \ref{fig_MMSEandRMMSE}.
It can be seen from both plots that the performance of BD is quite poor for systems with limited feedback.
Also, the achievable rate of MMSE performs close to that of MRT at low SNR region and that of BD at high SNR region.
However, MMSE experiences a rate loss after moderate SNR because interference from quantization errors degrades the performance.
Although WMMSE exhibits the best achievable rate when perfect CSIT is available, WMMSE also shows a rate degradation after SNR = 15 dB in limited feedback systems.
In contrast, the proposed designs demonstrate good performance as they can compensate for quantization errors by employing second-order statistics.
RMMSE outperforms conventional MRT, BD, and MMSE precoding schemes, and RWMMSE exhibits the best achievable rate with increased complexity.
By comparing Fig. \ref{fig_SR_10bits_824} and Fig. \ref{fig_SR_10bits_1628}, it can be observed that the gap between RMMSE and RWMMSE gets smaller as the number of users increases.
This can be attributed to two reasons.
An increased number of transmit antennas results in large quantization distortion.
Also, as the number of users grows, the accumulation of quantization distortions increases.
Thus, in order to counteract this effect and widen the gap, more feedback bits should be employed.

In Fig. \ref{fig_SR_VarB}, we exhibit the achievable rate with respect to different feedback bits.
One can notice from this figure that the achievable rate of all precoding schemes improves with the increasing number of feedback bits, while BD performs the worst achievable rate.
For SNR of 10 dB, MRT performs even better than BD.
MMSE and WMMSE outperform MRT if $B$ grows.
We can confirm that the proposed robust precoding designs perform better than conventional schemes.
For SNR = 10 dB with B = 10, RMMSE and RWMMSE can provide 7.1\% and 8.3\% sum-rate improvements over MMSE and WMMSE, respectively.
Also, for SNR = 30 dB with B = 10, we have 24.4 \% and 48.6 \% improvement for RMMSE and RWMMSE over MMSE and WMMSE, respectively.
Thus, we can confirm the superiority of the proposed RMMSE and RWMMSE precoding designs.

\begin{figure}[t]
\centering
\includegraphics[width=3.5in]{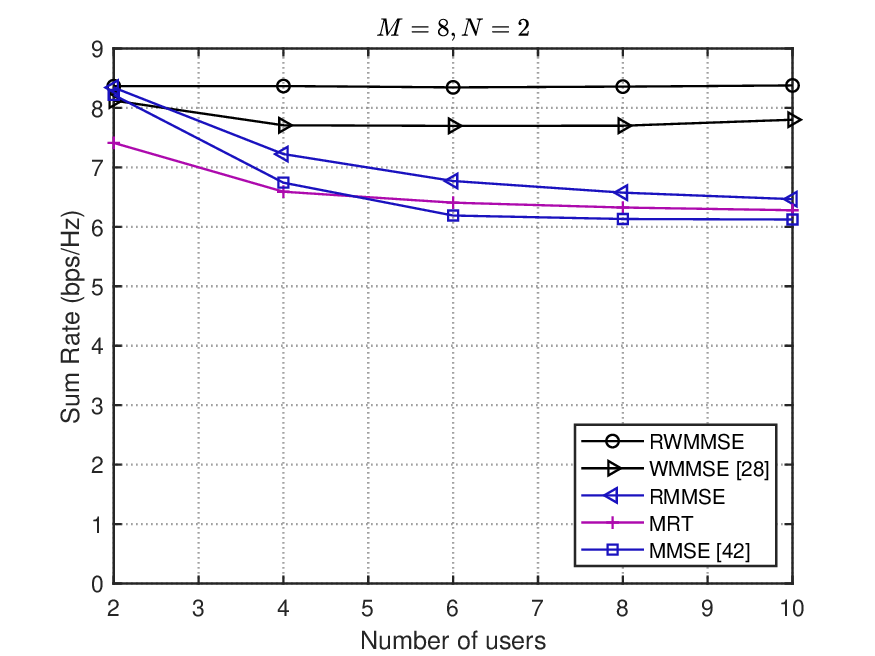}
\caption{Achievable rate of limited-feedback MU-MIMO systems with $B = 10$ and SNR = 10 dB.}
\label{fig_DiffK}
\end{figure}

Finally, the achievable rate with respect to the number of users is shown in Fig. \ref{fig_DiffK}.
Note that BD cannot be compared since it does not work in the overloaded case ($M<NK$).
From the plot, the achievable rate of non-iterative precoding schemes is degraded with increasing users.
On the contrary, the iterative precoding schemes exhibit a significant improvement.
This advantage arises from the power allocation capability of iterative precoding designs.
Specifically, non-iterative precoding schemes cannot allocate the power properly, which results in high interference at each user.
However, iterative precoding schemes allocate more power to the signal streams with less interference.
Thus, we can verify that RWMMSE works well in various system settings.

\section{Conclusion}
\label{Conclusion}
In this paper, we have analyzed the limitation of achievable rate in MU-MIMO systems with limited feedback.
To compensate for quantization errors, we have proposed new robust precoding designs by applying an approximation for the second-order statistics in terms of the quantized CSI and the number of feedback bits.
Based on the derived second-order statistics, we have extended traditional precoding to a robust case.
Then, we have presented a robust weighted MMSE precoding design with iterative solutions.
The numerical results have verified that our proposed schemes overcome the rate degradation at high SNR region and outperform conventional schemes in MU-MIMO systems with limited feedback.

\appendices
\section{Proof of Theorem \ref{theorem_achierate}}
\label{appendix_theorem_achierate}
For analysis simplicity, we set $\No$ to 1.
From Lemma \ref{lemma_ergodicequiv}, we can approximate the achievable rate of user $k$ through EVD as
\begin{equation*}
\begin{aligned}
R_k \approx & \log_2\Bigg| \brI_N + \bbE \llp \Hk^H \Pk \Pk^H \Hk\lrp \\ &\times \slp\sum\nolimits_{j \neq k}^{K} \bbE \llp \Hk^H \Pj \Pj^H \Hk \lrp + \In \srp^{-1} \Bigg| \\
= & \underbrace{\log_2\sll \brI_N + \rho \sum_{j =1}^{K} \bbE \llp \wtildeHk^H \barP_j \barP_j^H \wtildeHk \lrp\srl}_{R_{k,\text{all}}} \\
& - \underbrace{\log_2\sll \brI_N + \rho \sum_{j \neq k}^{K} \bbE \llp \wtildeHk^H \barP_j \barP_j^H \wtildeHk \lrp\srl}_{R_{k, \text{int}}},
\end{aligned}
\end{equation*}
where the last equation follows from the equal power allocation and $\bbE \llp \Lambdak \lrp = M \brI_N$.
As discussed in Section \ref{ProblemFormulation}, the MMSE precoding becomes close to the MRT at low SNR region (i.e., $\barP_j = \whatH_j, \forall j$), which makes $\whatHk$ and $\barP_j \slp j\neq k\srp$ independent.
Then, we calculate $R_{k,\text{int}}^\text{low}$ as
\begin{equation*}
\begin{aligned}
R_{k, \text{int}}^\text{low} & = \log_2\sll \brI_N + \rho \slp K-1 \srp \wtildeHk^H \bbE \llp \barP_j \barP_j^H \lrp \wtildeHk \srl \\
& = \log_2 \sll \brI_N + \rho \slp K-1 \srp \frac{N}{M}\brI_N \srl \\
& = N \log_2 \slp 1 + \rho \frac{K-1}{K} \srp.
\end{aligned}
\end{equation*}
Also, $R_{k, \text{all}}^\text{low}$ is rewritten as
\begin{equation*}
\begin{aligned}
R_{k, \text{all}}^\text{low}  = & \log_2 \left| \brI_N + \rho \slp K-1 \srp \wtildeHk^H \bbE \llp \barP_j \barP_j^H \lrp \wtildeHk \right. \\ &
\left. + \ \rho \bbE \llp \wtildeHk^H \barPk \barPk^H \wtildeHk \lrp \right|.
\end{aligned}
\end{equation*}
Recalling the channel decomposition model \eqref{chandecomp}, we have
\begin{equation*}
\begin{aligned}
\wtildeHk^H \barPk = \slp \whatHk \brX_k \brY_k + \brS_k \brZ_k \srp^H \whatHk = \brY_k^H \brX_k^H.
\end{aligned}
\end{equation*}
Then, it is obvious that $R_{k, \text{all}}^\text{low}$ is expressed as
\begin{equation*}
\begin{aligned}
R_{k, \text{all}}^\text{low}& = \log_2\sll \brI_N +  \rho \slp K-1 \srp \frac{N}{M}\brI_N + \rho \bbE \llp \brY_k^H \brY_k \lrp \srl \\
& =  \log_2\sll \brI_N +  \rho \frac{K-1}{K}\brI_N + \rho \slp1-\gamma \srp \brI_N \srl \\
& = N \log_2\slp 1 + \rho \slp 2 -\frac{1}{K} -\gamma \srp \srp.
\end{aligned}
\end{equation*}
Combining $R_{k, \text{int}}^\text{low}$ and $R_{k, \text{all}}^\text{low}$, the achievable rate of user $k$ at low SNR is approximated by
\begin{equation*}
\begin{aligned}
R_k^\text{low} = N \log_2 \slp 1+ \frac{\rho K \slp 1 - \gamma \srp}{\rho K - \rho + K} \srp.
\end{aligned}
\end{equation*}

At high SNR region, the MMSE precoding approaches the BD, which makes the $k$-th precoding matrix $\Pk$ orthogonal to $\whatH_j \slp j\neq k\srp$ and independent of $\whatHk$.
Under this criterion, we have
\begin{equation*}
\begin{aligned}
\bbE \llp \wtildeHk^H \barP_j \barP_j^H \wtildeHk \lrp & =  \bbE \llp \brZ_k^H \brS_k^H \barP_j \barP_j^H \brS_k \brZ_k \lrp \\
& = \bbE \llp \brZ_k^H \bbE \llp \brS_k^H \barP_j \barP_j^H \brS_k \lrp \brZ_k\lrp,
\end{aligned}
\end{equation*}
where we utilize the conditional expectation over $\brZ_k$.
Since both $\barP_k$ and $\brS_j \slp j\neq k\srp$ are independent and lie in the left nullspace of $\whatHk$, $\brS_k^H \barP_j \barP_j^H \brS_k$ follows a $\text{Beta}\slp N, M-2N \srp$ distribution, which has the expectation as $\bbE \llp \brS_k^H \barP_j \barP_j^H \brS_k \lrp = \frac{N}{M-N}\brI_N$ \cite{LFBD08}.
Therefore, $R_{k, \text{int}}^\text{high}$ and $R_{k, \text{all}}^\text{high}$ can be calculated as
\begin{equation*}
\begin{aligned}
R_{k, \text{int}}^\text{high} &  = \log_2\sll \brI_N + \rho \gamma \slp K-1 \srp \frac{N}{M-N} \brI_N \srl \\
& =  N \log_2\slp 1 + \rho \gamma \srp ,\\
R_{k, \text{all}}^\text{high} & = \log_2\sll \brI_N + \rho \gamma \slp K-1 \srp \frac{N}{M-N} \brI_N + \rho\frac{N}{M} \brI_N \srl \\
& = N \log_2\slp 1 + \rho \slp \frac{1}{K} + \gamma \srp \srp.
\end{aligned}
\end{equation*}
By combining $R_{k, \text{int}}^\text{high}$ and $R_{k, \text{all}}^\text{high}$, the achievable rate of the MMSE precoding at high SNR region is approximated by
\begin{equation*}
\begin{aligned}
R_{k}^\text{high} = N \log_2 \slp 1+ \frac{\rho}{K + \rho K \gamma} \srp.
\end{aligned}
\end{equation*}
Since each user has the same achievable rate, we omit the subscript for brevity.
$\hfill \blacksquare$

\section{Proof of Lemma \ref{lemma_equality}}
\label{appendix_lemma_equality}
First, we can see that $\brY^H \brX^H \brX \brY  = \brY^H \brY $.
Since $\brY$ consists of i.i.d. entries, $\bbE \llp \brY^H \brY \lrp$ becomes a diagonal matrix $\frac{1}{N}\sum_{n=1}^N \lambda_{n}^\bry \brI_N$. 
For the right-hand side of \eqref{lemma_eq}, we take the conditional expectation as \cite{RV02}
\begin{equation*}
\begin{aligned}
\bbE_{\brX,\brY} \llp \brX \brY \brY^H \brX^H \lrp = \bbE_{\brX} \llp \bbE_{\brY|\brX} \llp \brX \brY \brY^H \brX^H \lrp \lrp.
\end{aligned}
\end{equation*}
Also, since both $\brX$ and $\brY$ are independent of each other, and $\brY$ is a random matrix in terms of $\bbE \llp \brY \brY^H \lrp = \text{diag}\llp \bm{\lambda}^\bry \lrp$, it is evident that $\brX \bbE \llp \brY \brY^H \lrp \brX^H$ follows the Wishart distribution.
Then we have $\bbE_{\brX} \llp \bbE_{\brY|\brX} \llp \brX \brY \brY^H \brX^H \lrp \lrp$ equals $\frac{\text{tr}\slp\bbE \llp \brY \brY^H \lrp \srp}{N}\brI_N$.
Since $\frac{1}{N}\sum_{n=1}^N \lambda_{n}^\bry \brI_N=\frac{\text{tr}\slp\bbE \llp \brY \brY^H \lrp \srp}{N}\brI_N$, this completes the proof.
$\hfill \blacksquare$

\section{Proof of Theorem \ref{theorem_Rk}}
\label{appendix_theorem_Rk}
As both $\brR_{k,\mathbf{ZY}}$ and $\brR_{k,\mathbf{YZ}}$ have the same statistical property, we focus on $\brR_{k,\mathbf{ZY}}$ for brevity.
Let us take the singular value decomposition (SVD) to $\brZ_k \brY_k^H = \brJ_k \mathbf{\Omega}_k \brL_k^H$, where $\brJ_k$ and $\brL_k$ represent $N \times N$ unitary matrices, and $\mathbf{\Omega}_k=\text{diag}\llp \bm{\omega} \lrp$ is a diagonal matrix with $\bm{\omega}= \left[ \omega_{k,1},\cdots,\omega_{k,N}\right]^T$.
Defining $\brA_k=\brS_k \brJ_k$ and $\brB_k = \whatHk\brX_k\brL_k$, we can deduce $\brA_k^H\brB_k=\mathbf{0}$ and rewrite $\brR_{k,\mathbf{ZY}}$ as
\begin{equation*}
\begin{aligned}
\brR_{k,\mathbf{ZY}} & = M \bbE \llp \brA_k \mathbf{\Omega}_k \brB_k^H \lrp \\
& = M \bbE_{\brA_k,\brB_k}\llp \bbE_{\mathbf{\Omega}_k|\brA_k,\brB_k} \llp \brA_k \mathbf{\Omega}_k \brB_k^H \lrp\lrp \\
& = M \bbE\llp \brA_k \bbE \llp \mathbf{\Omega}_k \lrp \brB_k^H \lrp.
\end{aligned}
\end{equation*}
It can be seen that $\brR_{k,\mathbf{ZY}}$ approaches a zero matrix if the diagonal elements in $\bbE \llp \mathbf{\Omega}_k\lrp$ are small.
Thus, we focus on the condition that the elements in $\bbE \llp \mathbf{\Omega}_k\lrp$ tend to zero.

Now, we examine an upper bound of diagonal elements in $\bbE \llp \mathbf{\Omega}_k\lrp$.
$\omega_{k,n}$ in $\mathbf{\Omega}_k$ can be obtained by EVD as
\begin{equation*}
\begin{aligned}
\omega_{k,n} & = \sqrt{\text{eig}_n \slp \brY_k \brZ_k^H \brZ_k \brY_k^H  \srp} = \sqrt{\text{eig}_n \slp \brZ_k^H \brZ_k \brY_k^H \brY_k \srp},
\end{aligned}
\end{equation*}
where the equality follows from the fact that both $\brY_k \brZ_k^H \brZ_k \brY_k^H$ and $\brZ_k^H \brZ_k \brY_k^H \brY_k$ have the same eigenvalues.
Since $\brZ_k^H \brZ_k$ is a Hermitian matrix and $\brY_k^H\brY_k = \brI_N - \brZ_k^H\brZ_k$, $\omega_{k,n}$ can be rewritten as
\begin{equation*}
\begin{aligned}
\omega_{k,n} & = \sqrt{\text{eig}_n \slp \brZ_k^H \brZ_k \slp \brI_N - \brZ_k^H\brZ_k \srp \srp} \\
& = \sqrt{\text{eig}_n \slp \brZ_k^H \brZ_k -  \brZ_k^H \brZ_k \brZ_k^H\brZ_k \srp} \\
& = \sqrt{\text{eig}_n \slp \brZ_k^H \brZ_k \srp - \text{eig}_n \slp \brZ_k^H \brZ_k \srp^2}.
\end{aligned}
\end{equation*}
Since $\mathbf{\Omega}_k$ consists of unordered singular values, it follows $\bbE \llp \text{eig}_n \slp \brZ_k^H \brZ_k \srp \lrp = \gamma$.
Then, we have
\begin{equation*}
\begin{aligned}
\overline{\omega}_{k,n} =\bbE \llp \omega_{k,n} \lrp & = \bbE \llp \sqrt{\text{eig}_n \slp \brZ_k^H \brZ_k \srp - \text{eig}_n \slp \brZ_k^H \brZ_k \srp^2} \lrp \\
& \leq \sqrt{\bbE \llp \text{eig}_n \slp \brZ_k^H \brZ_k \srp - \text{eig}_n \slp \brZ_k^H \brZ_k \srp^2 \lrp} \\
& \leq \sqrt{ \gamma - \text{eig}_n \slp \bbE \llp \brZ_k^H \brZ_k \lrp \srp ^2} \\
& = \sqrt{\gamma \slp 1- \gamma \srp},
\end{aligned}
\end{equation*}
where Jensen's inequality is applied to the square root and square functions.
Since each diagonal element in $\bbE \llp \mathbf{\Omega}_k \lrp$ is upper bounded by $\sqrt{\gamma \slp 1- \gamma \srp}$, it is evident that both $\brR_{k,\mathbf{ZY}}$ and $\brR_{k,\mathbf{YZ}}$ are negligible if the quantization distortion $\gamma$ is small\footnote{$\sqrt{\gamma \slp 1- \gamma \srp}$ becomes small when the quantization distortion approaches 0 or 1. However, for a practical system, such as $M=8$, $N=2$, and $B=1$, $\gamma$ is typically no greater than 0.7. Therefore, $\sqrt{\gamma \slp 1- \gamma \srp}$ becomes negligible only with a small $\gamma$}.

As the quantization distortion is bounded by $\left[0,1\right]$, we also have $0\leq \overline{\omega}_{k,n}\leq 1$.
Then, we investigate $\bbE\llp \brA_k \bbE \llp \mathbf{\Omega}_k\lrp \brB_k^H \lrp$.
Since $\brS_k$ is a random semi-unitary matrix, each element in $\brA_k$ is a zero-mean complex random variable with independently distributed entries of variance $\frac{1}{M}$.
Define the $\slp m, n \srp$-th element of $\brA_k$ and $\brB_k$ as $a_{k, mn}$ and $b_{k, mn}$, respectively.
For a small $B$, the expectation of codewords is not zero, and thus the $\slp i, j \srp$-th element of $\bbE\llp \brA_k \bbE \llp \mathbf{\Omega}_k\lrp \brB_k^H \lrp$ can be rewritten as $\frac{1}{\sqrt{M}}\sum_{n=1}^N \overline{\omega}_{k, n} \bbE \llp \widetilde{a}_{k,in} b_{k,jn}^* \lrp$, where $\widetilde{a}_{k, in}$ is a complex number with unit variance.
Then, we have almost surely that \cite{LSNLP12}
\begin{equation*}
\begin{aligned}
\frac{1}{\sqrt{M}}\sum_{n=1}^N \overline{\omega}_{k, n} \bbE \llp \widetilde{a}_{k,in} b_{k,jn}^* \lrp \overset{M\rightarrow \infty}{\longrightarrow} 0.
\end{aligned}
\end{equation*}
Furthermore, if $B$ is not too small, the expectation of codewords approaches 0, and the variance of each element in $\brB_k$ gets close to $\frac{1}{M}$.
Then, we have almost surely that
\begin{equation*}
\begin{aligned}
\frac{1}{M}\sum_{i=1}^N \overline{\omega}_{k,i} \bbE \llp \widetilde{a}_{k, in} \widetilde{b}_{k, jn}^* \lrp \overset{M\rightarrow \infty}{\longrightarrow} 0,
\end{aligned}
\end{equation*}
where $\widetilde{b}_{k, jn}$ is a complex number with unit variance.
As a result, for a given number of feedback bits, increasing the number of transmit antennas also makes each element in $\brR_{k,\mathbf{YZ}}$ and $\brR_{k,\mathbf{ZY}}$ approach 0.

Thus, $\brR_k$ can be approximated as
\begin{equation*}
\begin{aligned}
\brR_k^o = \brR_k^c + \brR_k^n,
\end{aligned}
\end{equation*}
where $\brR_k^c = M \bbE \llp \whatHk \brX_k \brY_k \brY_k^H \brX_k^H \whatHk^H  \lrp$ and $\brR_k^n = M \bbE \llp  \brS_k \brZ_k \brZ_k^H  \brS_k^H \lrp$.
According to Lemma \ref{lemma_equality} with $\bbE \llp \brY_k^H \brY_k \lrp = \slp 1- \gamma \srp \brI_N$, we have $\brR_k^c = \slp1-\gamma \srp \whatHk \whatHk^H$.
As $\brR_k^n$ accounts for the random components, $\brR_k^n$ is important in the derivation of $\brR_k^o$.
Since $\brS_k$ is isotropically distributed in the left nullspace of $\whatHk$, we can decompose $\brS_k$ as $\brS_k = \brQ_k \brG_k$, where $\brQ_k \in \CD^{M \times N}$ has the same subspace as $\brS_k$ and $\brG_k \in \CD^{N \times N}$ is a unitary matrix (see Lemma 1 in \cite{CIMU09}).
Thus, reusing Lemma \ref{lemma_equality} with $\bbE \llp \brZ_k^H \brZ_k \lrp = \gamma \brI_N$ results in $\brR_k^n=\bbE \llp  \brQ_k \brG_k \brZ_k \brZ_k^H  \brG_k^H \brQ_k^H\lrp = \gamma \bbE \llp \brQ_k \brQ_k^H \lrp$.

To further investigate the covariance matrix of $\brQ_k$, $\brQ_k$ is decomposed as $\brQ_k=\widehat{\brU}_k \brV_k$ where $\widehat{\brU}_k \in \CD^{M \times \slp M-N\srp}$ consists of the orthonormal basis of the left nullspace of $\whatHk$, and $\brV_k \in \CD^{\slp M-N \srp \times N}$ is a random semi-unitary matrix.
In addition, since $[\whatHk, \widehat{\brU}_k ]$ forms an orthonormal matrix, we utilize $\widehat{\brU}\widehat{\brU}^H = \brI_M - \whatHk \whatHk^H$ to derive the covariance of $\brQ_k$ as
\begin{equation*}
\begin{aligned}
\bbE \llp \brQ_k \brQ_k^H \lrp & = \bbE \llp \widehat{\brU}_k \brV_k \brV_k^H \widehat{\brU}_k^H \lrp \\
& =  \widehat{\brU}_k \bbE \llp \brV_k \brV_k^H \lrp \widehat{\brU}_k^H \\
& = \frac{N}{M-N}\slp \brI_M -  \whatHk \whatHk^H\srp,
\end{aligned}
\end{equation*}
where the final equality comes from $\bbE \llp \brV_k \brV_k^H \lrp = \frac{N}{M-N} \brI_{M-N}$.
Then, we obtain an approximation of $\brR_k$ by combining $\brR_k^c$ and $\brR_k^n$.
This completes the proof.
$\hfill \blacksquare$

\section{Proof of Lemma \ref{lemma_chandecomp}}
\label{appendix_lemma_chandecomp}
From Theorem \ref{theorem_Rk}, we can see that the approximated second-order statistic $\brR_k^o$ in \eqref{chancovmtx} is composed of two independent terms which are related to quantized CSI and an identity matrix, respectively.
This motivates us to apply the Karhunen-Loeve representation \cite{KLR13} separately on each of these two terms.
For the first term, the basis is obtained from the quantized CSI $\whatHk$.
As for the second term, we generate a random zero-mean semi-unitary matrix $\brO_k \in \CD^{M\times N}$.
Each element of $\brO_k$ is an i.i.d. complex Gaussian random variable with variance of $\frac{1}{M}$.
The channel decomposition is then given as
\begin{equation}
\label{KLchan}
\begin{aligned}
\Hk \approx \delta_k \whatHk+ \eta_k \brO_k,
\end{aligned}
\end{equation}
where $\delta_k$ satisfies $\bbE\llp\delta_k\lrp = \sqrt{M-\frac{M^2\gamma}{M-N}}$.
Also, we can check that the covariance matrix of $\brO_k$ is $\bbE \llp \brO_k \brO_k^H \lrp = \frac{N}{M} \brI_M$.
To make sure that the second-order statistic of \eqref{KLchan} is identical to $\brR_k^o$, we have $\bbE \llp \eta_k \lrp = \sqrt{ \frac{M^2 \gamma}{M-N}}$.
Since each element in $\brO_k$ has a zero mean, the conditional expectation of $\Hk$ is approximated by
\begin{equation*}
\begin{aligned}
\bbE _{\brH|\whatH} \llp\brH_k\lrp \approx \delta \whatHk.
\end{aligned}
\end{equation*}
This completes the proof.
$\hfill \blacksquare$

\section{Proof of Lemma \ref{lemma_expectation}}
\label{appendix_lemma_expectation}
Defining $\brx_k = \left[x_{k1}, \cdots, x_{kM} \right]^T$ and $\bry_k = \left[y_{k1}, \cdots, y_{kM} \right]^T$ as the $k$-th column vector of $\brX$ and $\brY$, respectively.
The element-wise multiplication of $\brK\triangleq \bbE \llp \brY^H \brX \brY \lrp$ is calculated by
\begin{equation*}
\begin{aligned}
\brK & = 
\bbE \llp 
\left[ \begin{array}{ccc}
\sum_{i=1}^M \bry_1^H \brx_i y_{1i} & \ldots & \sum_{i=1}^M \bry_1^H \brx_i y_{Ni} \\
\vdots & \ddots & \vdots \\
\sum_{i=1}^M \bry_N^H \brx_i y_{1i} & \ldots & \sum_{i=1}^M \bry_N^H \brx_i y_{Ni}
\end{array} \right]
\lrp \\
& = \left[ \begin{array}{ccc}
\sum_{i=1}^M \bbE \llp | y_{1i}|^2 \lrp x_{ii}  & \ldots & 0 \\
\vdots & \ddots & \vdots  \\
0 & \ldots & \sum_{i=1}^M \bbE \llp | y_{Ni}|^2 \lrp x_{ii} 
\end{array} \right].
\end{aligned}
\end{equation*}
Since the entries in $\brY$ are generated in terms of $\sigma_{m,n}^2, \forall m, n$, we have
\begin{equation*}
\begin{aligned}
\brK & = \text{diag} \llp 
\left[ \begin{array}{ccc}
\sigma_{11}^2 & \ldots & \sigma_{1M}^2 \\
\vdots & \ddots & \vdots  \\
\sigma_{N1}^2 & \ldots & \sigma_{NM}^2 
\end{array} \right] 
\left[ \begin{array}{ccc}
x_{11} \\
\vdots  \\
x_{MM}
\end{array} \right] \lrp.
\end{aligned}
\end{equation*}
$\hfill \blacksquare$

\ifCLASSOPTIONcaptionsoff
  \newpage
\fi

\begin{IEEEbiography}[{\includegraphics[width=1in,height=1.2in,clip,keepaspectratio]{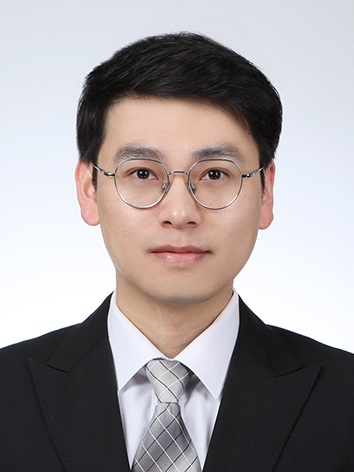}}]{Wentao Zhou}
(S'20) received the B.S. degree from Guilin University of Electronic Technology, Guilin, China, in 2017, and the M.S. degree from Soongsil University, Seoul, South Korea, in 2020.
He is currently working towards the Ph.D. degree with the School of Electrical Engineering at Korea University, Seoul, South Korea.

His current research interests include multiple antenna communications, limited feedback techniques, and rate-splitting multiple access.
\end{IEEEbiography}

\begin{IEEEbiography}[{\includegraphics[width=1in,height=1.25in,clip,keepaspectratio]{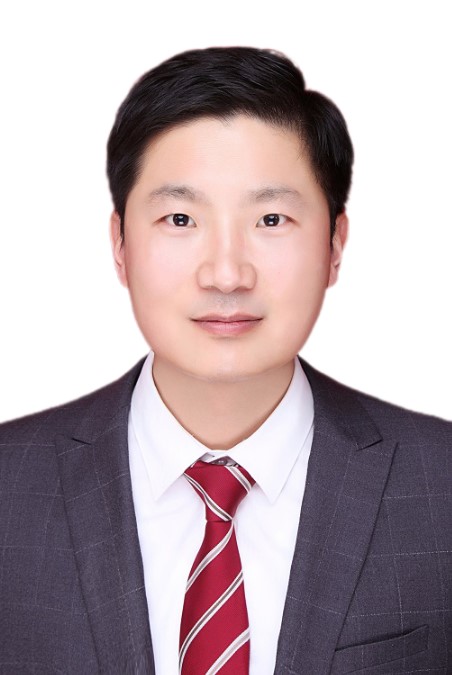}}]{Di Zhang}
(S'13, M'17, SM'19) currently is an Associate Professor at Zhengzhou University, he is also a Visiting Scholar of Korea University. He is serving as an area editor of KSII Transactions on Internet and Information Systems, has served as the guest editor of \textsc{IEEE Wireless Communications} and \textsc{IEEE Network}. He received the First Prize Award for Science and Technology Progress of Henan Province in 2023, the First Prize Award for Science and Technology Achievements from Henan Department of Education in 2023, and the ITU Young Author Recognition in 2019. 

His research interests are the wireless communications and networking, especially the short packet communications and its applications.
\end{IEEEbiography}

\begin{IEEEbiography}[{\includegraphics[width=1in,height=1.25in,clip,keepaspectratio]{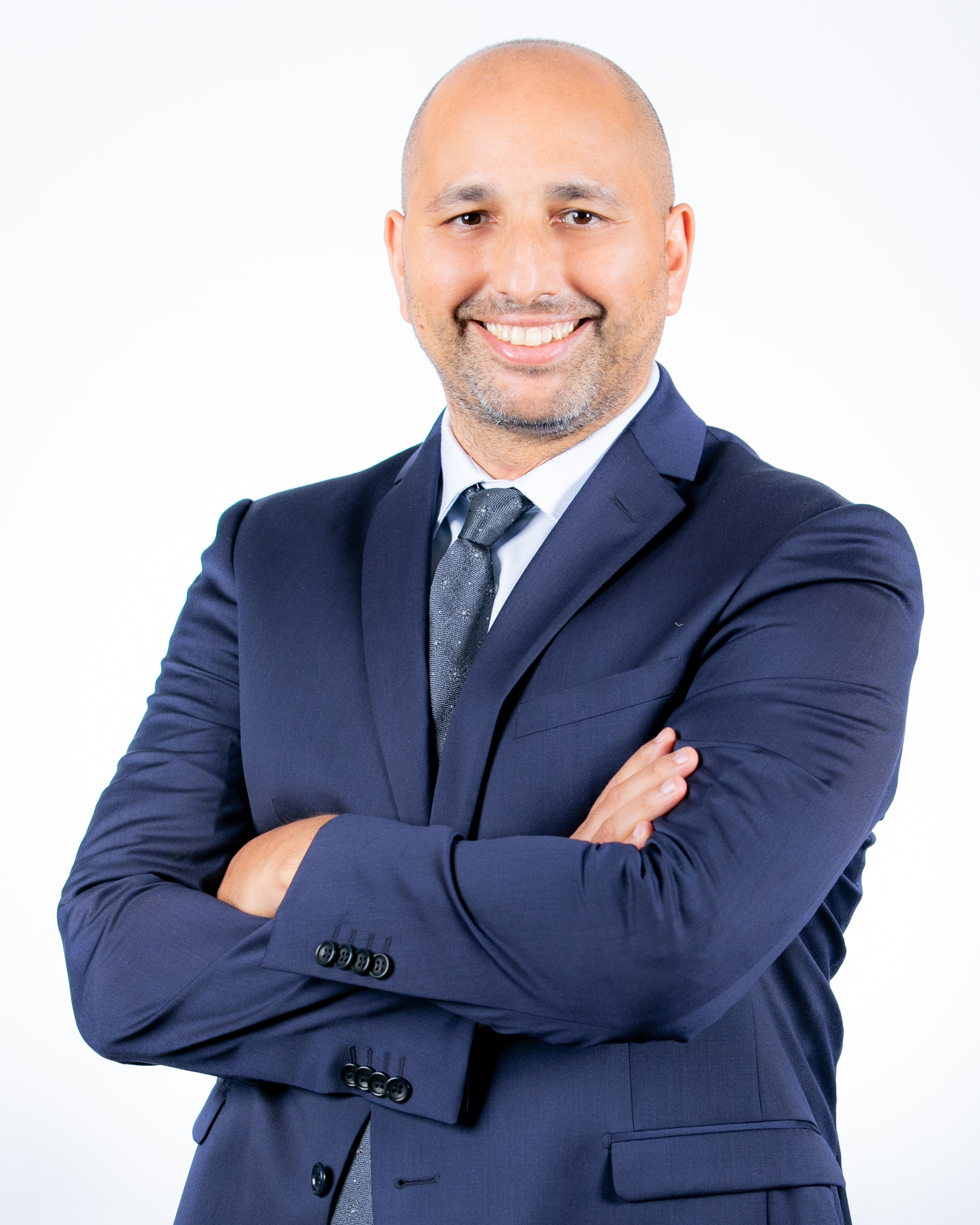}}]{M\'{e}rouane Debbah}
is Professor at Khalifa University of Science and Technology in Abu Dhabi and founding Director of the KU 6G Research Center. He is affiliated as an Adjunct Professor with the School of Electrical Engineering at Korea University. He is a frequent keynote speaker at international events in the field of telecommunication and AI. His research has been lying at the interface of fundamental mathematics, algorithms, statistics, information and communication sciences with a special focus on random matrix theory and learning algorithms. In the Communication field, he has been at the heart of the development of small cells (4G), Massive MIMO (5G) and Large Intelligent Surfaces (6G) technologies. In the AI field, he is known for his work on Large Language Models, distributed AI systems for networks and semantic communications. He received multiple prestigious distinctions, prizes and best paper awards (more than 40 IEEE best paper awards) for his research contributions. He is an IEEE Fellow, a WWRF Fellow, a Eurasip Fellow, an AAIA Fellow, an Institut Louis Bachelier Fellow and a Membre \'{e}m\'{e}rite SEE.
\end{IEEEbiography}

\begin{IEEEbiography}[{\includegraphics[width=1in,height=1.25in,clip,keepaspectratio]{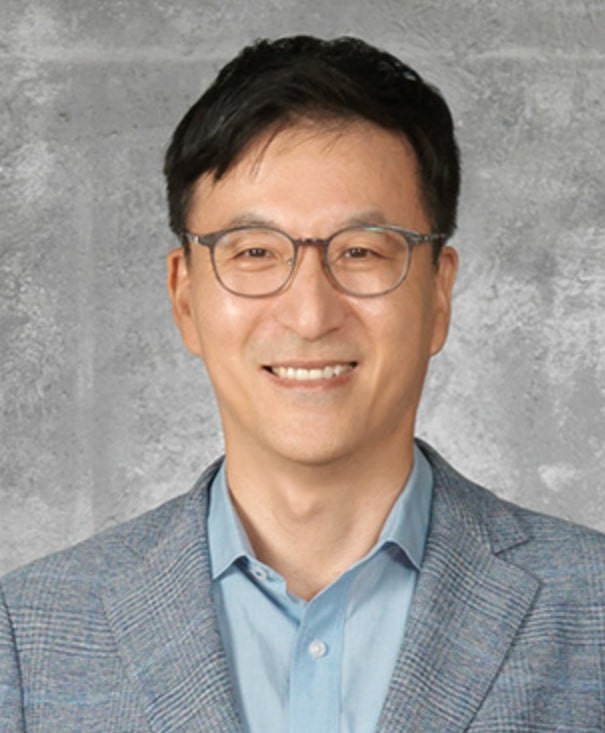}}]{Inkyu Lee}
(Fellow, IEEE) received the B.S. degree (Hons.) in control and instrumentation engineering from Seoul National University, Seoul, South Korea, in 1990, and the M.S. and Ph.D. degrees in electrical engineering from Stanford University, Stanford, CA, USA, in 1992 and 1995, respectively. From 1995 to 2002, he was a member of the Technical Staff with Bell Laboratories, Lucent Technologies, Murray Hill, NJ, USA, where he studied high-speed wireless system designs. Since 2002, he has been with Korea University, Seoul, where he is currently a Professor with the School of Electrical Engineering. He has also served as the Department Head of the School of Electrical Engineering, Korea University, from 2019 to 2021. In 2009, he was a Visiting Professor with the University of Southern California, Los Angeles, CA, USA. He has authored or coauthored more than 220 journal articles in IEEE publications and holds 30 U.S. patents granted or pending. His research interests include digital communications and signal processing techniques applied for next-generation wireless systems. He was a recipient of the IT Young Engineer Award from the IEEE/IEEK Joint Award in 2006, the Best Paper Award from the IEEE Vehicular Technology Conference in 2009, the Best Research Award from the Korean Institute of Communications and Information Sciences in 2011, the Best Paper Award from the IEEE International Symposium on Intelligent Signal Processing and Communication Systems in 2013, the Best Young Engineer Award from the National Academy of Engineering of Korea in 2013, and the Korea Engineering Award from the National Research Foundation of Korea in 2017. He served as an Associate Editor for the \textsc{IEEE Transactions on Communications} from 2001 to 2011 and the \textsc{IEEE Transactions on Wireless Communications} from 2007 to 2011. He was a Chief Guest Editor of the \textsc{IEEE Journal on Selected Areas in Communications} Special Issue on “4G wireless systems” in 2006. He was a TPC co-chair for IEEE International Conference on Communications (ICC) in 2022. He also serves as the Co-Editor-in-Chief for the Journal of Communications and Networks. He was elected as a member of the National Academy of Engineering of Korea in 2015. He is the Director of “Augmented Cognition Meta-Communication” ERC Research Center awarded from the National Research Foundation of Korea. He is a Distinguished Lecturer of IEEE.
\end{IEEEbiography}


\begin{thebibliography}{1}

\bibitem{VTC2024}
W. Zhou, D. Zhang, M. Debbah, and I. Lee, ``Robust MMSE precoding for limited-feedback multiuser MIMO systems,'' in \textit{Proc. IEEE 99th Veh. Technol. Conf.,} Singapore,  Jun. 2024, to appear.

\bibitem{OVERVIEW08}
 D. J. Love, R. W. Heath, V. K. Lau, D. Gesbert, B. D. Rao, and M. Andrews, ``An overview of limited feedback in wireless communication systems,'' \textit{IEEE J. Sel. Areas Commun.}, vol. 26, no. 8, pp. 1341--1365, Oct. 2008.

\bibitem{LFHP15}
A. Alkhateeb, G. Leus, and R. W. Heath, ``Limited feedback hybrid precoding for multi-user millimeter wave systems,'' \textit{IEEE Trans. Wireless Commun.}, vol. 14, no. 11, pp. 6481--6494, Nov. 2015.

\bibitem{RSLNR16}
H. Shen, W. Xu, A. Swindlehurst, and C. Zhao, ``Transmitter optimization for per-antenna power constrained multi-antenna downlinks: An SLNR maximization methodology,'' \textit{IEEE Trans. Signal Process.}, vol. 64, no. 10, pp. 2712--2725, May 2016.

\bibitem{RSMISO16}
H. Joudeh and B. Clerckx, ``Sum-rate maximization for linearly precoded downlink multiuser MISO systems with partial CSIT: A rates-plitting approach,'' \textit{IEEE Trans. Commun.}, vol. 64, no. 11, pp. 4847--4861, Nov. 2016.

\bibitem{RBDEH19}
Z. Zhu, S. Huang, Z. Chu, F. Zhou, D. Zhang, and I. Lee, ``Robust designs of beamforming and power splitting for distributed antenna systems with wireless energy harvesting,'' \textit{IEEE Systems Journal}, vol. 13, no. 1, pp. 30--41, Mar. 2019.

\bibitem{GPIP20}
J. Choi, N. Lee, S. N. Hong, and G. Caire, ``Joint user selection, power allocation, and precoding design with imperfect CSIT for multi-cell MU-MIMO downlink systems,'' \textit{IEEE Trans. Wireless Commun.}, vol. 19, no. 1, pp. 162--176, Jan. 2020.

\bibitem{RBD21}
Z. Zhu, N. Wang, W. Hao, Z. Wang, and I. Lee, ``Robust beamforming designs in secure MIMO SWIPT IoT networks with a nonlinear channel model,'' \textit{IEEE Internet Things J.}, vol. 8, no. 3, pp. 1702–1715, Feb. 2021.

\bibitem{RLTD21}
K. P. Rajput, Y. Verma, N. K. D. Venkategowda, A. K. Jagannatham, and P. K. Varshney, ``Robust linear transceiver designs for vector parameter estimation in MIMO wireless sensor networks under CSI uncertainty,'' \textit{IEEE Trans. Veh. Technol.}, vol. 70, no. 8, pp.7347--7362, Aug. 2021.

\bibitem{RSMAMO21}
O. Dizdar, Y. Mao, and B. Clerckx, ``Rate-splitting multiple access to mitigate the curse of mobility in (massive) MIMO networks,'' \textit{IEEE Trans. Commun.}, vol. 69, no. 10, pp. 6765--6780, Oct. 2021.

\bibitem{RJPD23}
M. Kazemi, Ç. Göken, and T. M. Duman, ``Robust joint precoding/combining design for multiuser MIMO systems with calibration errors,'' \textit{IEEE Trans. Wireless Commun.}, vol. 22, no. 8, pp. 5157--5169, Aug. 2023.

\bibitem{BOHI21}
H. Shen, W. Xu, S. Gong, C. Zhao, and D. W. K. Ng, ``Beamforming optimization for IRS-aided communications with transceiver hardware impairments,'' \textit{IEEE Trans. Commun.}, vol. 69, no. 2, pp. 1214--1227, Feb. 2021.

\bibitem{MIMIHI23}
J. Wang, S. Gong, Q. Wu, and S. Ma, ``RIS-aided MIMO systems with hardware impairments: robust beamforming design and analysis,'' \textit{IEEE Trans. Wireless Commun.}, vol. 22, no. 10, pp. 6914--6929, Oct. 2023.

\bibitem{LFBD08}
N. Ravindran and N. Jindal, ``Limited feedback-based block diagonalization for the MIMO broadcast channel,'' \textit{IEEE J. Sel. Areas Commun.}, vol. 26, no. 8, pp. 1473--1482, Oct. 2008.

\bibitem{OFRIL16}
J. Park, N. Lee, J. G. Andrews, and R. W. Heath, ``On the optimal feedback rate in interference-limited multi-antenna cellular systems,'' \textit{IEEE Trans. Wireless Commun.}, vol. 15, no. 8, pp. 5748--5762, Aug. 2016.

\bibitem{OAMG16}
M. Min, Y.-S. Jeon, and G.-H. Im, ``On achievable multiplexing gain of BD in MIMO broadcast channels with limited feedback,'' \textit{IEEE Trans. Wireless Commum.}, vol. 15, no. 2, pp. 871--885, Feb. 2016.

\bibitem{OARUS17}
M. Min, Y.-S. Jeon, and G.-H. Im, ``On achievable rate of user selection for MIMO broadcast channels with limited feedback,'' \textit{IEEE Trans. Commun.}, vol. 65, no. 1, pp. 122--135, Jan. 2017.

\bibitem{OOCSI19}
M. Min, ``On the optimal CSI feedback rate of downlink MIMO systems using multiple receive antennas in cellular networks,'' \textit{IEEE Wireless Commun. Lett.}, vol. 8, no. 6, pp. 1612--1616, Jul. 2019

\bibitem{ACDC08}
N. Jindal, ``Antenna combining for the MIMO downlink channel,'' \textit{IEEE Trans. Wireless Commun.}, vol. 7, no. 10, pp. 3834--3844, Oct. 2008.

\bibitem{SQBC13}
S. Schwarz and M. Rupp, ``Subspace quantization based combining for limited feedback block-diagonalization,'' \textit{IEEE Trans. Wireless Commum.}, vol. 12, no. 11, pp. 5868--5879, Nov. 2013.

\bibitem{MEAR15}
S. Schwarz and M. Rupp, ``Maximum expected achievable rate combining for limited feedback block-diagonalization,'' in \textit{Proc. IEEE Int. Conf. Acoust. Speech Signal Process.}, Brisbane, Qld., Australia, Apr. 2015, pp. 3093--3097.

\bibitem{CQLF14}
S.-H. Moon, S.-R. Lee, J.-S. Kim, and I. Lee, ``Channel quantization for block diagonalization with limited feedback in multiuser MIMO downlink channels,'' \textit{J. Commun. Netw.}, vol. 16, no. 1, pp. 1--9, Feb. 2014.

\bibitem{PFIA20}
N. Garg, A. K. Jagannatham, G. Sharma, and T. Ratnarajah, ``Precoder feedback schemes for robust interference alignment with bounded CSI uncertainty,'' \textit{IEEE Trans. Signal Inf. Process. Netw.}, vol. 6, pp. 407--425, May 2020.

\bibitem{NOMAQO20}
X. Zou, M. Ganji, H. Jafarkhani, ``Downlink asynchronous non-orthogonal multiple access with quantizer optimization,'' \textit{IEEE Wireless Commun. Lett.}, vol. 9, no. 10, pp. 1606--1610, Oct. 2020.

\bibitem{DLCSI23}
W. Kim, Y. Ahn, J. Kim and B. Shim, ``Towards deep learning-aided wireless channel estimation and channel state information feedback for 6G,'' \textit{J. Commun. Netw.}, vol. 25, no. 1, pp. 61--75, Feb. 2023.

\bibitem{RZF05}
C. B. Peel, B. M. Hochwald, and A. L. Swindlehurst, ``A vector perturbation technique for near-capacity multiantenna multiuser communication---Part I: Channel inversion and regularization,'' \textit{IEEE Trans. Commun.}, vol. 53, no. 1, pp. 195--202, Jan. 2005.

\bibitem{SLNR07}
M. Sadek, A. Tarighat, and A. H. Sayed, ``A leakage-based precoding scheme for downlink multi-user MIMO channels,'' \textit{IEEE Trans. Wireless Commun.}, vol. 6, no. 5, pp. 1711–1721, May 2007.

\bibitem{WMMSE08}
S. S. Christensen, R. Agarwal, E. Carvalho, and J. M. Cioffi, ``Weighted sum-rate maximization using weighted MMSE for MIMO-BC beamforming design,'' \textit{IEEE Trans. Wireless Commun.}, vol. 7, no. 12, pp. 4792--4799, Dec. 2008.

\bibitem{RSLNR18}
T. X. Tran and K. C. Teh, ``Spectral and energy efficiency analysis for SLNR precoding in massive MIMO systems with imperfect CSI,'' \textit{IEEE Trans. Wireless Commun.}, vol. 17, no. 6, pp. 4017–4027, Jun. 2018.

\bibitem{RWMMSE13}
R. Fritzsche and G. P. Fettweis, ``Robust sum rate maximization in the multi-cell MU-MIMO downlink,'' in \textit{Proc. IEEE Wireless Commun. Netw. Conf. (WCNC)}, Shanghai, China, Apr. 2013, pp. 3180--3184.

\bibitem{RSMA22}
A. Mishra, Y. Mao, O. Dizdar, and B. Clerckx, ``Rate-splitting multiple access for downlink multiuser MIMO: Precoding optimization and PHY-layer design,'' \textit{IEEE Trans. Commun.}, vol. 70, no. 2, pp. 874--890, Feb. 2022.

\bibitem{PACE13}
S. K. Mohammed and E. G. Larsson, ``Per-antenna constant envelope precoding for large multi-user MIMO systems,'' \textit{IEEE Trans. Commun.}, vol. 61, no. 2, pp. 1059–1071, Mar. 2013.

\bibitem{GLSE19}
A. Bereyhi, M. A. Sedaghat, R. R. Müller, and G. Fischer, ``GLSE precoders for massive MIMO systems: Analysis and applications,'' \textit{IEEE Trans. Wireless Commun.}, vol. 18, no. 9, pp. 4450–4465, Sep. 2019.

\bibitem{RHP21}
Z. Luo, L. Zhao, L. Tonghui, H. Liu, and R. Zhang, ‘''Robust hybrid precoding/combining designs for full-duplex millimeter wave relay systems,'' \textit{IEEE Trans. Veh. Technol.}, vol. 70, no. 9, pp. 9577–9582, Sep. 2021.

\bibitem{RHT23}
P. Maity, K. P. Rajput, S. Srivastava, N. K. D. Venkategowda, A. K. Jagannatham, and L. Hanzo, ``Robust hybrid transceiver designs for linear decentralized estimation in mmWave MIMO IoT networks in the face of imperfect CSI,'' \textit{IEEE Internet Things J.}, vol. 10, no. 20, pp. 18125--18139, Oct. 2023.

\bibitem{RMMSE08}
A. D. Dabbagh and D. J. Love, ``Multiple antenna MMSE based downlink precoding with quantized feedback or channel mismatch,'' \textit{IEEE Trans. Commun.}, vol. 56, no. 11, pp. 1859--1868, Nov. 2008.

\bibitem{RMMSE09}
C. Zhang, W. Xu, and M. Chen, ``Robust MMSE beamforming for multiuser MISO systems with limited feedback,'' \textit{IEEE Signal Process. Lett.}, vol. 16, no. 7, pp. 588--591, Jul. 2009.

\bibitem{EEOLF23}
T. Wu and Y. Zou, ``Energy efficiency optimization in adaptive transmit antenna selection systems with limited feedback,'' \textit{IEEE Internet Things J.}, vol. 10, no. 2, pp. 1248--1258, Jan. 2023.

\bibitem{LRFB16}
X. Yang and A. L. Swindlehurst, ``Limited rate feedback in a MIMO wiretap channel with a cooperative jammer,'' \textit{IEEE Trans. Wireless Commum.}, vol. 64, no. 18, pp. 4695--4706, Sep. 2016.
 
\bibitem{QALF21}
M. AlaaEldin, E. Alsusa, and K. G. Seddik, ``Quantized vs. analog channel feedback for FDD massive MIMO systems with multiple-antenna users,'' in \textit{Proc. IEEE Pers., Indoor Mobile Radio Commun. (PIMRC)}, Helsinki, Finland, Sep. 2021, pp. 684--690.

\bibitem{JAPLD22}
Z. Tang, L. Sun, D. Niyato, Y. Zhang, and A. Liu, ``Jammer-assisted secure precoding and feedback design for MIMO IoT networks,'' \textit{IEEE Internet Things J.}, vol. 9, no. 14, pp. 12241--12257, Jul. 2022.

\bibitem{MMSE05}
M. Joham, W. Utschick, and J. A. Utschick, ``Linear transmit processing in MIMO communications systems,'' \textit{IEEE Trans. Signal Process.}, vol. 53, no. 8, pp. 2700--2712, Aug. 2005.

\bibitem{LSNLP12}
S. Wagner, R. Couillet, M. Debbah, and D. T. M. Slock, ``Large system analysis of linear precoding in correlated MISO broadcast channels under limited feedback,'' \textit{IEEE Trans. Inf. Theory}, vol. 58, no. 7, pp. 4509--4537, Jul. 2012.

\bibitem{PSUP14}
Q. Zhang, S. Jin, K.-K. Wong, H. Zhu, and M. Matthaiou, ``Power scaling of uplink massive MIMO systems with arbitrary-rank channel means,'' \textit{IEEE J. Sel. Topics Signal Process.}, vol. 8, no. 5, pp. 966--981, Oct. 2014.

\bibitem{MatAna16}
J. R. Schott, \textit{Matrix Analysis for Statistics}. New York: Wiley, 2016.

\bibitem{CO04}
S. Boyd and L. Vandenberghe, \textit{Convex Optimization}. Cambridge, U.K.: Cambridge Univ. Press, 2004.

\bibitem{NP99}
D. Bertsekas, \textit{Nonlinear Programming}, 2nd ed. Belmont, MA: Athena Scientific, 1999.

\bibitem{FLOP07}
R. Hunger, “Floating point operations in matrix-vector calculus,” Technische Universität München, Munchen, Germany, Tech. Rep. TUM-LNSTR-05-5, Sep. 2007.

\bibitem{RV02}
A. Papoulis and S. U. Pillai, \textit{Probability, Random Variables, and Stochastic Processes}, 4th ed. New York, NY, USA: McGraw-Hill, 2002.

\bibitem{CIMU09}
H. Sung, S.-R. Lee, and I. Lee, ``Generalized channel inversion methods for multiuser MIMO systems,'' \textit{IEEE Trans. Commun.}, vol. 57, no. 11, pp. 3489--3499, Nov. 2009.

\bibitem{KLR13}
A. Adhikary, J. Nam, J.-Y. Ahn, and G. Caire, ``Joint spatial division and multiplexing---The large-scale array regime,'' \textit{IEEE Trans. Inf. Theory}, vol. 59, no. 10, pp. 6441--6463, Oct. 2013.

\end{thebibliography}
\end{document}